\documentclass[10pt]{iopart}

\usepackage{etoolbox}
\usepackage{hyperref}
\usepackage[caption=false]{subfig}

\expandafter\let\csname equation*\endcsname=\relax
\expandafter\let\csname endequation*\endcsname=\relax
\usepackage{color}
\usepackage{amsmath}
\usepackage{amssymb}
\usepackage{enumerate}
\usepackage{graphicx}
\usepackage{mathbbol}
\usepackage{amsfonts}
\usepackage{url}

\newcommand{\nn}{\nonumber}

\newcommand{\bea}{\begin{eqnarray}}
\newcommand{\eea}{\end{eqnarray}}
\newcommand{\beq}{\begin{equation}}
\newcommand{\eeq}{\end{equation}}

\usepackage[utf8]{inputenc}
\usepackage{amsmath}
\usepackage{hyperref}
\usepackage{graphicx}
\usepackage{amsfonts}
\usepackage{amsthm}
\usepackage{cases}
\usepackage{bm}
\usepackage{amssymb}

\usepackage{color}
\definecolor{Blue}{rgb}{0.00, 0.00, 1.00}
\definecolor{Red}{rgb}{1.00, 0.00, 0.00}

\hypersetup{
    colorlinks=true,       
    linkcolor=blue,          
    citecolor=magenta,        
    filecolor=red,      
    urlcolor=cyan           
}
\newcommand{\be}{\begin{equation}}
\newcommand{\ee}{\end{equation}}

\newcommand{\beqn}{\begin{eqnarray}}
\newcommand{\eeqn}{\end{eqnarray}}

\DeclareMathOperator{\erf}{erf}
\DeclareMathOperator{\sign}{sign}

\newcommand{\moy}[1]{\ensuremath{\langle #1 \rangle}}

\newcommand{\var}[1]{{\rm Var}\left(#1\right)}

\makeatletter
\renewcommand\@appendixstar{\@@par
 \ifnumbysec 
 \@addtoreset{table}{section}
 \@addtoreset{figure}{section}\fi
 \setcounter{section}{0}
 \setcounter{subsection}{0}
 \setcounter{subsubsection}{0}
 \setcounter{equation}{0}
 \setcounter{figure}{0}
 \setcounter{table}{0}
 \def\thesection{\Alph{section}} 
 \def\theequation{\ifnumbysec
      \Alph{section}.\arabic{equation}\else
      \Alph{section}\arabic{equation}\fi}
 \def\thetabLe{\ifnumbysec
      \Alph{section}\arabic{tabLe}\else
      A\arabic{tabLe}\fi}
 \def\thefigure{\ifnumbysec
      \Alph{section}\arabic{figure}\else
      A\arabic{figure}\fi}}
\makeatother

\def\Xint#1{\mathchoice
   {\XXint\displaystyle\textstyle{#1}}%
   {\XXint\textstyle\scriptstyle{#1}}%
   {\XXint\scriptstyle\scriptscriptstyle{#1}}%
   {\XXint\scriptscriptstyle\scriptscriptstyle{#1}}%
   \!\int}
\def\XXint#1#2#3{{\setbox0=\hbox{$#1{#2#3}{\int}$}
     \vcenter{\hbox{$#2#3$}}\kern-.5\wd0}}

\def\dashint{\Xint-}

\begin{document}
\title[Brownian coincidences]{Distribution of Brownian coincidences}

\author{Alexandre Krajenbrink}
\address{Laboratoire de Physique de l’Ecole Normale Sup\'erieure,\\ PSL University, CNRS, Sorbonne Universit\'es,\\
 24 rue Lhomond, 75231 Paris Cedex 05, France}

\author{Bertrand Lacroix-A-Chez-Toine}
\address{LPTMS, CNRS, Univ. Paris-Sud, Universit\'e Paris-Saclay, 91405 Orsay, France}

\author{Pierre Le Doussal}
\address{Laboratoire de Physique de l’Ecole Normale Sup\'erieure,\\ PSL University CNRS, Sorbonne Universit\'es,\\
 24 rue Lhomond, 75231 Paris Cedex 05, France}

\begin{abstract}
We study the probability distribution, $P_N(T)$, of the coincidence time $T$, i.e. the total local time of all pairwise coincidences
of $N$ independent Brownian walkers. We consider in details two geometries: Brownian motions 
all starting from $0$, and Brownian bridges. Using a Feynman-Ka\v c representation for the moment generating function of this coincidence time, we map this problem onto some observables in three related models
(i) the propagator of the Lieb Liniger model of quantum particles with pairwise delta function interactions (ii) the moments of the partition function of a directed polymer in a random medium (iii) the exponential moments of the solution of the Kardar-Parisi-Zhang equation. Using these mappings, we obtain closed formulae for the probability distribution of the coincidence time, its tails and some of its moments. Its asymptotics at large and small coincidence time are also obtained for arbitrary fixed endpoints. The universal large $T$ tail, $P_N(T) \sim \exp(- 3 T^2/(N^3-N))$ is obtained, and is independent of the geometry. 
We investigate the large deviations in the limit of a large number of walkers through a Coulomb gas approach.
Some of our analytical results are compared with numerical simulations.
\end{abstract}

\maketitle

{\hypersetup{linkcolor=black}
	\tableofcontents
}
\newpage

\section{Introduction and main results}

\subsection{Introduction}

In this article, we study independent and identical diffusive Brownian processes. We want to characterise in details the statistics of the time spent by identical diffusing particles in the vicinity 
of each other,
 which we call the coincidence time. This problem is relevant 
for instance
 to analyse reaction networks of the type
\be
{\rm A}\;\;+\;\;{\rm A}\;\;\to\;\;{\rm B}\;.
\ee
Indeed, for chemical species to react, they first have to encounter each other and then overcome the potential barriers. This takes a finite amount of time and this coincidence time plays a crucial role in the kinetics of the reaction. The behaviour of this coincidence time is highly dependent on the spatial dimension, as it is clear that the higher the dimension, the harder it is for diffusing particles to encounter one another. We limit our study of this problem to the case of dimension $d=1$, where the number of 
encounters
 is maximum.
We model the identical diffusing particles via independent one dimensional Brownian motions $x_i(\tau)$ on the time interval $\tau \in [0,t]$
\be\label{Langevin}
\dot x_{i}(\tau)=\eta_i(\tau)\;\;{\rm with}\;\;\moy{\eta_i(\tau)}=0\;\;\moy{\eta_i(\tau)\eta_j(\tau')}=2D\delta_{i,j}\delta(\tau-\tau')\;.
\ee
where below we choose units such that $D=1$. 
We consider the case where the $N$ diffusing particles are emitted at $t=0$ from a single point-like source $x_i(0)=x_{0}$ for $i =1, \dots , N$.
We define the total time spent by these particles in a close vicinity of length $\ell$ to each other as
\be
T_N(\ell;t)=\sum_{i \neq  j}^N \int_0^{t}\Theta\left[\frac{\ell}{2}-|x_i(\tau)-x_j(\tau)|\right]\rmd \tau\;,
\ee
where $\Theta(x)$ is the Heaviside step-function. We are particularly interested in the limit where the length $\ell$ is small compared to all the other scales of the problem and therefore define in the limit $\ell\to 0$,
\be \label{def}
{\cal T}_N(t)=\lim_{\ell\to 0}\frac{1}{\ell}T_N(\ell;t)=\sum_{i \neq j}^N \int_0^{t}\delta[x_i(\tau)-x_j(\tau)]\rmd \tau\;.
\ee
We will refer to this observable as the coincidence time of $N$ diffusing particles, \textit{i.e.} the amount of time that these independent particles spend crossing each other \footnote{
for calculational 
convenience below, each pair appears twice in the sum in our definition \eqref{def}.
}.
Note however that ${\cal T}_N(t)$ does not have the dimension of a time, 
which is a usual property of
the local time of a stochastic process \cite{revuz2013continuous, Bo89, Pit99}.
Using the Brownian scaling, the process has a trivial rescaling to the time interval $\tau \in [0,1]$, and we obtain the equality
\be
{\cal T}_N(t=1)={\cal T}_N=\frac{{\cal T}_N(t)}{\sqrt{t}}\;.
\ee
We will see in the following that the $N$-dependence is however highly non-trivial. The case $N=2$ can be solved quite easily by considering the process $z(\tau)=\frac{1}{\sqrt{2}}(x_1(\tau)-x_2(\tau))$ which is also a simple diffusive process of same diffusion coefficient $D=1$,
\be
\dot{z}(\tau)=\xi(\tau)\;,\;\;{\rm with}\;\;\moy{\xi(\tau)}=0\;\;\moy{\xi(\tau)\xi(\tau')}=2D\delta(\tau-\tau')\;.
\ee
The coincidence time ${\cal T}_2$ of the two Brownian particles can be expressed in terms of the local time $L_t(x)$ of the process $z(\tau)$ defined as
\be
L_t(x)=\int_{0}^{t}\rmd \tau \delta(z(\tau)-x)\;.
\ee
Setting $x=0$, we obtain
\begin{align}\label{T_loc_T_oc}
L_t(0)&=\int_{0}^{t}\rmd \tau \delta(z(\tau))=\int_{0}^{t}\rmd \tau \delta\left(\frac{1}{\sqrt{2}}(x_1(\tau)-x_2(\tau))\right)\\
&=\sqrt{2}\int_{0}^{t}\rmd \tau \delta\left(x_1(\tau)-x_2(\tau)\right)=\frac{{\cal T}_2(t)}{\sqrt{2}}\;.\nn
\end{align}
The joint PDF of the local time $L_t(0)=L$ and the final position $z(t)=x_f$ was obtained by Borodin et al. \cite{Bo89} and exploited by Pitman \cite{Pit99} and reads 

\be\label{p_j_T_x_f}
{\cal P}_{\rm joint}(L,x_f)=\frac{1}{2} \sqrt{\frac{1}{D \pi t^3}}(|x_f|+2D L)e^{-\frac{(|x_f|+2D L)^2}{4D t}}\;.
\ee
We set $t=1$ and remind that $D=1$ in our convention. Using now the identities $L_1(0)={\cal T}_2/\sqrt{2}$ together with $z(1)=(x_1(1)-x_2(1))/\sqrt{2}$, we obtain the joint PDF of the coincidence time ${\cal T}_2=T$ and the final algebraic distance between the diffusive particles $d=x_1(1)-x_2(1)$ as
\be\label{P_joint_T_d}
P_{\rm joint}(T,d)=\frac{1}{2}{\cal P}_{\rm joint}\left(\frac{T}{\sqrt{2}},\frac{d}{\sqrt{2}}\right)=\frac{(|d|+2T)}{2 \sqrt{8\pi}}e^{-\frac{(|d|+2T)^2}{8}}\;.
\ee 
However, as soon as $N=3$, one cannot define independent processes for which our coincidence time is a simple observable.\\

\begin{figure}[h!]
\centering
\includegraphics[width=0.49\textwidth]{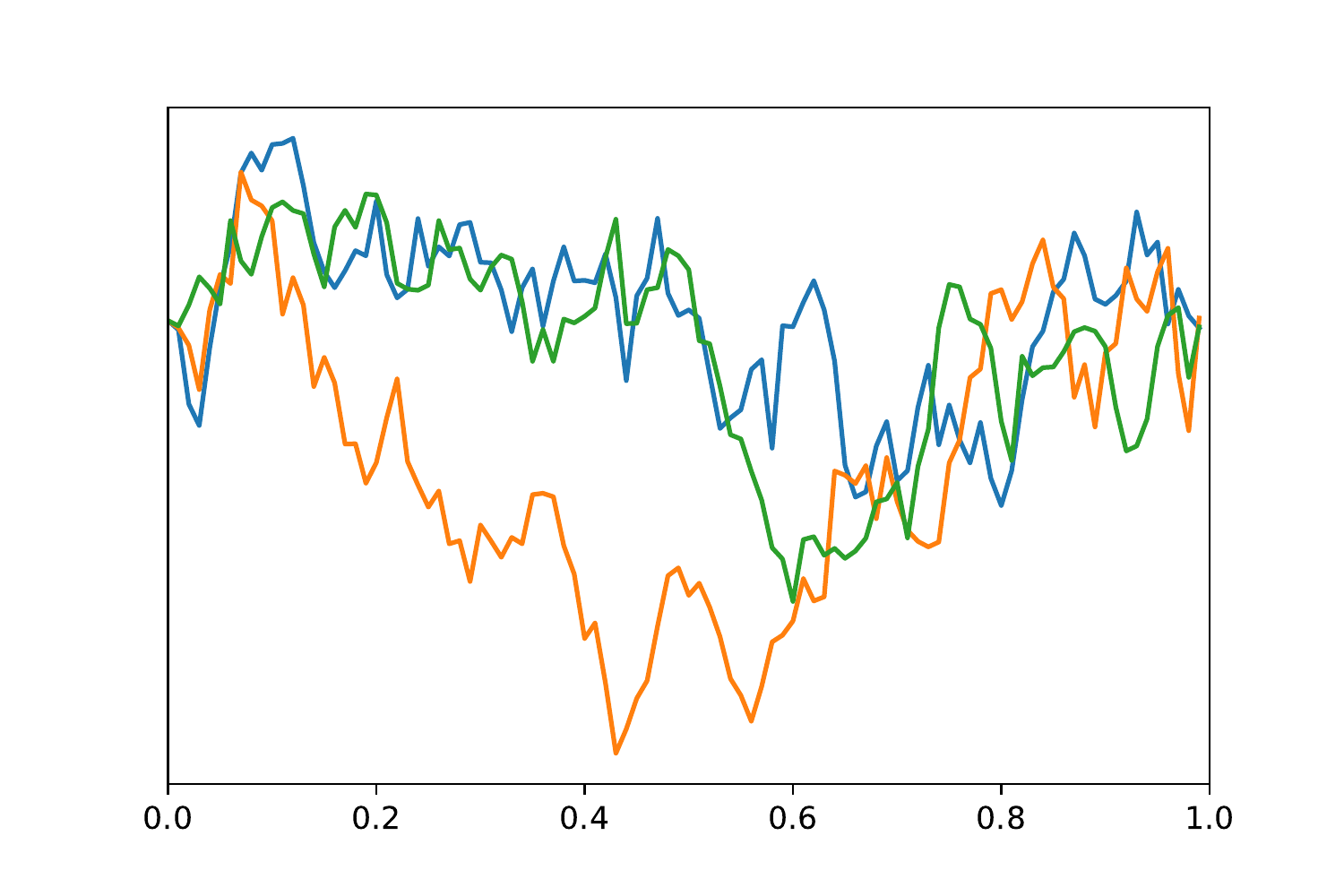}
\includegraphics[width=0.49\textwidth]{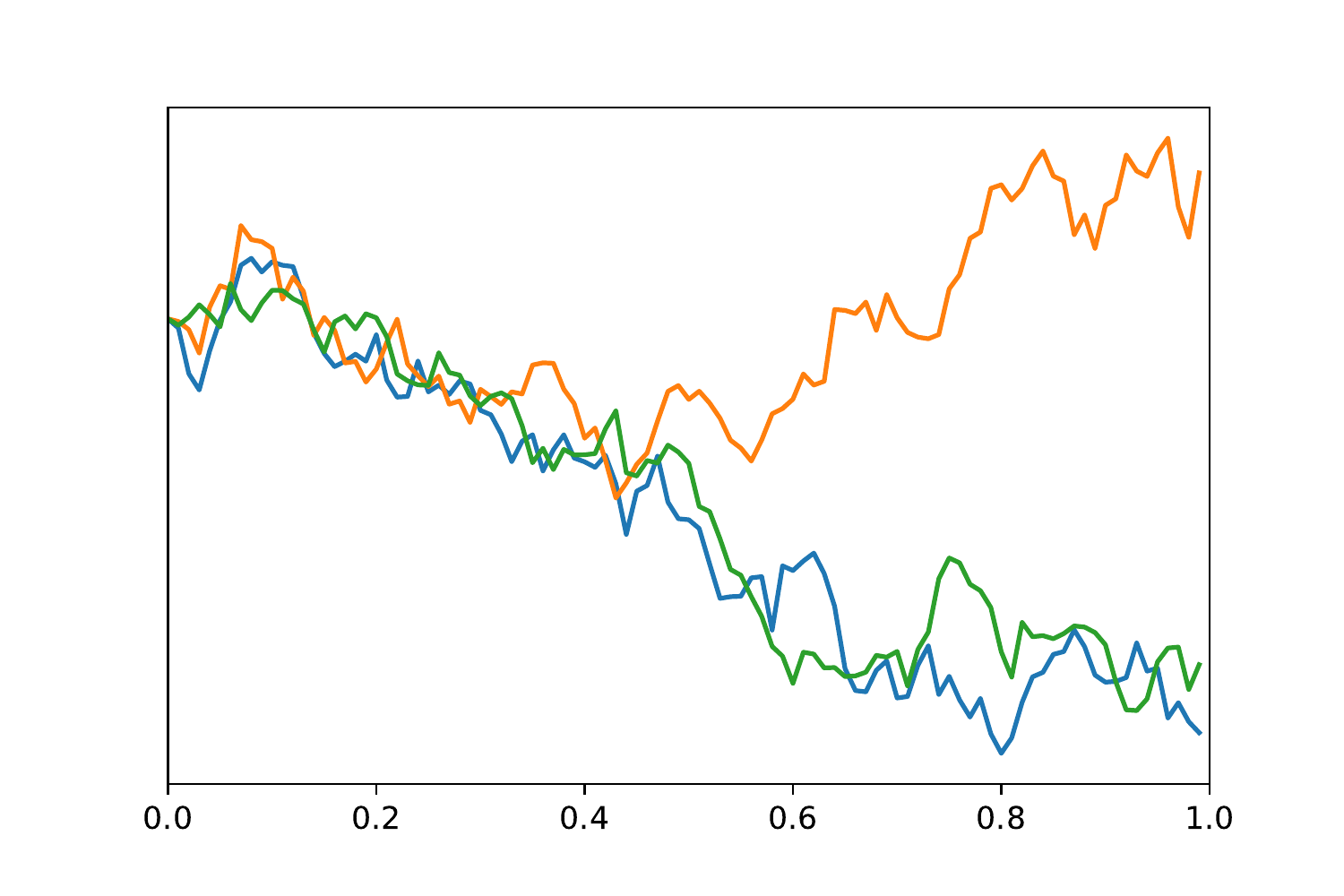}
\caption{Plot of a simulation of $N=3$ diffusive particles for a diffusion coefficient $D=1$ starting from ${\bf x}_0=(0,0,0)$. {\bf Left}: Brownian bridges with endpoints in ${\bf y}=(0,0,0)$. {\bf Right}: Brownian motions with arbitrary final points.
}\label{plot_bm_bb}
\end{figure}

Here we will study the Probability Distribution Function (PDF) of ${\cal T}_N$ for two cases, Brownian motions and
Brownian bridges. In the first case (Brownian motion) one defines the Moment Generating Function (MGF) $\moy{e^{-c {\cal T}_N(t)}}_{{\bf x}_0}$, where $\moy{\dots}_{{\bf x}_0}$ denotes the expectation value with respect to the PDF of ${\cal T}_N(t)$ for given initial conditions ${\bf x}(\tau=0)={\bf x}_0=(x_{0},\dots,x_{0})$.
This MGF can be expressed in the Feynman-Ka\v c framework as an $N$-dimensional path integral
\begin{align}
\moy{e^{-c {\cal T}_N(t)}}_{{\bf x}_0}&=\int_{\mathbb{R}^N} \rmd {\bf y} Z_N({\bf y},t|{\bf x}_0;c)\;,\;\;{\rm with}\;,\label{MGF_flat}\\
Z_N({\bf y},t|{\bf x}_0;c)&=\int_{{\bf x}(0)={\bf x}_0}^{{\bf x}(t)={\bf y}}{\cal D} {\bf x}(\tau)~ e^{-\displaystyle \int_0^{t}\left[\sum_{i=1}^N \frac{\dot{x}_i(\tau)^2}{4}+2 c \sum_{i<j}^N\delta[x_i(\tau)-x_j(\tau)]\right]\rmd \tau}\;,\label{LL_prop}
\end{align}
where ${\bf x}(\tau)=(x_1(\tau),x_2(\tau),\cdots,x_N(\tau))$. In this expression we wrote the sum $\sum_{i\neq j}$ as $2\sum_{i<j}$ to represent the coincidence as a delta interaction between each pair of Brownian motions. As the final point ${\bf y}$ is arbitrary, we integrated here over all the possible realisations. This situation corresponds to the right panel in Fig. \ref{plot_bm_bb}. In the second case (Brownian bridges)
we define the MGF as the following expectation value with fixed initial and final positions 
\be
\label{eq:mgf_general}
\moy{e^{-c {\cal T}_N(t)}}_{{\bf x}_0,{\bf y}}= \frac{Z_N({\bf y},t|{\bf x}_0;c)}{Z_N({\bf y},t|{\bf x}_0;c=0)}
\ee
where $Z_N({\bf y},t|{\bf x}_0;c=0)=(4 \pi t)^{-N/2} e^{- ({\bf y}-{\bf x}_0)^2/(4 t)}$ is the free propagator. This situation corresponds to the left panel in Fig. \ref{plot_bm_bb}.

The function $Z_N({\bf y},t|{\bf x}_0;c)$ can be interpreted as the imaginary time $N$-body propagator of the bosonic Lieb-Liniger Hamiltonian \cite{LL63}
\be\label{H_LL}
-\partial_{t}Z_N=\hat {\cal H}_N(c) Z_N\;\;{\rm with}\;\;\hat {\cal H}_N(c)=\sum_{i=1}^N {\hat p}_i^2 +2 c\sum_{i<j}^N\delta(\hat x_i-\hat x_j)\;.
\ee
We therefore see that this MGF, eq.~\eqref{eq:mgf_general}, for non-interacting diffusing classical particles is mapped onto the propagator  of a quantum problem of identical bosons with a contact interaction. This problem is also related to other well-studied models: the directed polymer in a random potential, equivalent to the Kardar-Parisi-Zhang 
(KPZ)
 stochastic growth equation.
Indeed, consider an elastic polymer of length $t$ and fixed endpoints $x_0$ and $y$,
in a random potential $\xi(x,t)$. Its partition function reads\\
\be
{\cal Z}(y,x_0,t)=\int_{{x}(0)={x}_0}^{{x}(t)={y}}{\cal D} {x}(\tau) {e}^{- \displaystyle \int_0^{t}\left[\frac{\dot{x}(\tau)^2}{4}+ \xi(x(\tau),\tau)\right]\rmd \tau}\;.
\ee
One may then compute the $N$-point correlation function of the partition function, averaged with respect to the random potential (denoted as $\overline{\cdots}$) and obtain
\be\label{rep_av}
\overline{\prod_{i=1}^N {\cal Z}(y_i,x_{0},t)}=\int_{{\bf x}(0)={\bf x}_0}^{{\bf x}(t)={\bf y}}{\cal D} {\bf x}(\tau)\overline{e^{-\displaystyle \int_0^{t}\sum_{i=1}^N \left[\frac{\dot{x}_i(\tau)^2}{4}+ \xi(x_i(\tau),\tau)\right]\rmd \tau}}\;.
\ee

Taking now $\xi(x,t)$ as a Gaussian white noise of variance $\bar c>0$,
\be
\overline{\xi(x,t)}=0\;\;{\rm and}\;\;\overline{\xi(x,t)\xi(x',t')}=2 \bar c \delta(x-x')\delta(t-t')\;,
\ee
the replica average in Eq. \eqref{rep_av} reduces to the Lieb-Liniger real time propagator in Eq. \eqref{LL_prop} \cite{kardareplica,bb-00,CDR10} with the value of the parameter $c=-\bar c$, and satisfies the same evolution equation \eqref{H_LL}\footnote{The constraint $i \neq j$ in \eqref{LL_prop} originates from the It\^o prescription
in the (Feynman-Ka\v c) stochastic heat equation satistied by ${\cal Z}$, equivalently in the
definition of the path integral. 
}. The directed polymer partition function ${\cal Z}(y,x_0,t)$ is also equal to the droplet solution of the KPZ equation, through a Cole-Hopf transformation,
\be
\log {\cal Z}(y,x_0,t)=h(y,x_0,t)\;,\;\;{\rm where}\;\;\partial_t h=\partial_{y}^2 h+ (\partial_y h)^2+\sqrt{2 \bar c} ~\xi(y,t)\;.
\ee
Both the directed polymer and the KPZ 
equation correspond to the choice $c = - \bar c<0$, \textit{i.e.} to the attractive 
Lieb-Liniger model. \\

The moments \eqref{rep_av} of the directed polymer problem, which
correspond to exponential moments for the KPZ height field,
 have been studied extensively recently and
exact formulae have been obtained for the droplet solution at arbitrary time \cite{CDR10,dotsenko,BorodinMacdo}.
The MGF of our diffusion problem, 
for
$N$ independent Brownian bridges, 
and
 negative value of the parameter $c = - \bar c$ is thus related to
these moments as follows
\be \label{formula} 
\moy{e^{ \bar c {\cal T}_N(t)}}_{{\bf 0},{\bf 0}} = \frac{\overline{{\cal Z}(0,0,t)^N}}{\mathcal Z_0(0,0,t)^N} 
\ee
where $\mathcal Z_0(x,0,t)=\frac{1}{\sqrt{4 \pi t}} e^{-x^2/(4 t)}$ is the free Brownian propagator (for $c=0$).
Explicit expressions of these moments were derived from the Bethe ansatz. 
Note that all initial and final points here are set to $0$.\\

In this paper, we obtain complementary exact formula for the MGF for positive value $c>0$, i.e. 
for the Laplace transform of the coincidence time $\moy{e^{ - c {\cal T}_N(t)}}$.
We study in particular the case of $N$ independent Brownian bridges 
denoted $\moy{e^{ - c {\cal T}_N(t)}}_{\rm BB} \equiv \moy{e^{-c {\cal T}_N(t)}}_{{\bf 0},{\bf 0}}$, and
 the case of $N$ Brownian motions starting from the same point, $x_0=0$, 
with 
free final points,
 denoted $\moy{e^{ - c {\cal T}_N(t)}}_{\rm BM}\equiv \moy{e^{ - c {\cal T}_N(t)}}_{{\bf 0}}$.
We also use the formula \eqref{formula} for negative $c=-\bar c<0$ to obtain the large $T$ asymptotics of
the PDF of ${\cal T}_N(t)$. For that 
particular asymptotic behavior
we obtain some more results for arbitrary fixed final points,
using a formula for $\moy{e^{-c {\cal T}_N(t)}}_{{\bf 0},{\bf y}}$. We now summarize our main
results.

\subsection{Summary of the main results}
~~~
\textbf{Brownian bridges} We obtain an exact formula for the PDF of the rescaled coincidence time ${\cal T}_N$ for $N$ Brownian bridges as 
\begin{align}
P_{N,{\rm BB}}(T)
&=\partial_T^{N+1} \int_{\mathbb{R}_+^{N}} \rmd {\bf r}\; \Theta\left(T-\sum_\ell r_\ell\right) \det_{1\leqslant i,j\leqslant N}\left(e^{-\frac{(r_i-r_{j})^2}{4}}\right)\label{PDF_P_BB_res}
\end{align}
For $N=2,3$, we obtain explicit formulae for $P_{N,{\rm BB}}(T)$ respectively in Eq. \eqref{P_2_BB_ana} and \eqref{P_3_BB_ana}.
For general $N$, we extract the small and large $T$ asymptotic behaviours as
\be
P_{N,{\rm BB}}(T)=\begin{cases}
\displaystyle 2^{- \frac{N(N-1)}{2}} \frac{G(N+2)}{\Gamma(N(N-1))}T^{N(N-1)-1}+\mathcal{O}(T^{N(N-1)+1})&\;,\;\;T\to 0\;,\\
&\\
\displaystyle \frac{N!2^{N-1}\pi^{\frac{N}{2}-1}}{N^{3/2}}\alpha_N^{\frac{N}{2}} H_{N-1}\left(\sqrt{\alpha_N}T\right)  e^{-\alpha_N T^2}+\mathcal{O}(e^{-\beta_N T^2}) &\;,\;\;T\to\infty\;,
\end{cases}
\ee 
where $G(n)=\prod_{k=1}^{n-2}k!$ is the Barnes-G function, $\Gamma$ is the usual Gamma function, $H_p(x)=e^{x^2}(-\partial_x)^p e^{-x^2}$ is the Hermite polynomial of degree $p$ and the exponential factors are
\be\label{a_b}
\alpha_N=\frac{3}{(N-1)N(N+1)}\;,\;\;\beta_N=\frac{3}{(N-2)(N-1)N}\;.
\ee
We also obtain the mean coincidence time in Eq. \eqref{mean_BB} and the variance in Eq. \eqref{var_BB}.\\

\textbf{Brownian motions} We obtain an exact formula for the MGF of the rescaled coincidence time ${\cal T}_N$ for $N$ 
Brownian motions (i.e. with free final points),
 which depend on the parity of $N$. It is given in Eq. \eqref{MGF_BM_even} for even values of $N$ and in Eq. \eqref{MGF_BM_odd} for odd values.
For $N=2,3$, we obtain explicit formulae for the PDF $P_{N,{\rm BM}}(T)$,
\be
P_{2,{\rm BM}}(T)=\sqrt{\frac{2}{\pi}} e^{- \frac{T^2}{2}}\;,\;\;P_{3,{\rm BM}}(T)=\sqrt{\frac{2}{\pi}} e^{- \frac{T^2}{2}} (e^{\frac{3}{8} T^2} -1)\;.
\ee
For arbitrary values of $N$, we obtained the expression of the PDF as:\\
\textbf{$N$ odd}
\be
\begin{split}
&P_{N,{\rm BM}}(T)=\\
&N\partial_T^{N+1} \int_{\mathbb{R}_+^{N}} \rmd {\bf r}\; \Theta\left(T-\sum_{\ell=1}^N r_\ell\right)\prod_{1\leqslant k <\ell \leqslant N} \, {\rm sign}(r_k-r_\ell)\;  \underset{1\leqslant k,\ell\leqslant N-1}{\rm Pf}\left[\, \erf\left(\frac{r_{k}-r_{\ell}}{\sqrt{2}}\right)\right]\label{PDF_P_BM_odd_intro}
\end{split}
\ee
\textbf{$N$ even}
\be
\begin{split}
&P_{N,{\rm BM}}(T)=\\
& \partial_T^{N+1} \int_{\mathbb{R}_+^{N}} \rmd {\bf r}\; \Theta\left(T-\sum_{\ell=1}^N r_\ell\right)\prod_{1\leqslant k <\ell \leqslant N} \, {\rm sign}(r_k-r_\ell)\;  \underset{1\leqslant k,\ell\leqslant N}{\rm Pf}\left[\, \erf\left(\frac{r_{k}-r_{\ell}}{\sqrt{2}}\right)\right]\label{PDF_P_BM_even_intro}
\end{split}
\ee
We further extracted the small and large $T$ behaviours as
\be
P_{N,{\rm BM}}(T)=\begin{cases}
\displaystyle I_N  T^{\frac{N(N-1)}{2}-1}+\mathcal{O}(T^{\frac{N(N-1)}{2}+1})&\;,\;\;T\to 0\;,\\
&\\
\displaystyle 2^{N-1}\sqrt{\frac{3}{\pi N(N^2-1)}}  e^{-\alpha_N T^2}+\mathcal{O}(e^{-\beta_N T^2}) &\;,\;\;T\to\infty\;,
\end{cases}
\ee 
where $I_N$ is given 
in Eq.~\eqref{INbis}. The coefficients 
$\alpha_N$ and $\beta_N$ are the same as for the Brownian bridge in Eq.~\eqref{a_b}. We also obtain the mean coincidence time in Eq.~\eqref{mean_BM} and the variance in Eq.~\eqref{var_BM}.\\

Finally, we extended our study to $N$ Brownian walkers with arbitrary fixed endpoints. The small $T$ asymptotics of the PDF for arbitrary final points ${\bf y}$ (all initial points being at $0$) is given in \eqref{Py}.
The large $T$ asymptotics of the PDF for any choice of initial and final points is displayed in \eqref{anyXY}.\\

A trivial consequence of our work is the PDF $P_N^c(T)$ of the coincidence time for $N$ Brownian walkers {\it interacting} pairwise via
a delta interaction of strength $c$ (of any sign). It is given in terms of the PDF for non-interacting Brownian $P_N(T)$ calculated here, simply as
\be
P_N^c(T) = \frac{e^{-c T} P_N(T)}{\int_0^{+\infty} \rmd T' e^{-c T'} P_N(T')}
\ee

The remaining of the paper is organised as follows. Section \ref{BB}, is dedicated to the results for the case of Brownian bridges. In subsection \ref{LL_BA}, 
we describe how the Bethe ansatz applied to the 
Lieb-Liniger model allows to obtain a formula for the MGF.
 In subsection \ref{BB_PDF}, we derive the exact formula in Eq. \eqref{PDF_P_BB_res} for the PDF 
of the coincidence time ${\cal T}_N=T$
 for arbitrary $N$. In subsection \ref{BB_23}, we analyse this formula for $N=2,3$. In subsection \ref{BB_asy}, we derive the asymptotic behaviours for $T\to 0$ and $T\to \infty$ for arbitrary $N$. In subsection \ref{BB_mean_var}, we obtain the mean and variance for arbitrary $N$. Finally, in subsection \ref{BB_CG}, we analyse the large $N$ limit via Coulomb gas techniques. In Section \ref{sec:BM}, we extend all previous results - except the Coulomb gas - to the Brownian motions in the same fashion and derive the exact PDF for arbitrary $N$ Brownian motions given in Eqs.~\eqref{PDF_P_BM_even_intro} and \eqref{PDF_P_BM_odd_intro}. 
In Section \ref{sec:coincidence}, we study  some
 properties of the coincidence time for $N$ Brownian with arbitrary fixed endpoints. We obtain the exact expression of its
 PDF for $N=2$ and $N=3$, the small $T$ asymptotics for any $N$ and the large $T$ asymptotics for arbitrary initial and final points.
The latter is obtained using the properties of the ground state of the Lieb-Liniger model.

\section{Coincidence  time  for Brownian bridges and KPZ equation with droplet initial conditions}\label{BB}

\subsection{Bethe ansatz solution of the Lieb-Liniger model for droplet initial conditions}\label{LL_BA}
\label{subsec:Bethe}

The Lieb-Liniger Hamiltonian in Eq. \eqref{H_LL} is an exactly solvable model using the Bethe ansatz. 
Here we need only the symmetric eigenstates of $\hat {\cal H}_N(c)$, which in the repulsive case $c>0$,  are single particle 
states. The eigen-energies $E_{\bf k}$ of this system defined on
a circle of radius $L$ form, in the large $L$ limit, a continuum indexed by 
$N$  real momenta $\lbrace k_i \rbrace $'s such that
\begin{align}
&
\hat {\cal H}_N(c) |\Psi^c_{\bf k}\rangle =E_{\bf k}|\Psi^c_{\bf k}\rangle\;,
\;\;{\rm with}\;\;E_{\bf k}={\bf k}^2=\sum_{i=1}^N k_i^2\;,\\
&\langle {\bf x}|\Psi_{\bf k}^c\rangle=\frac{C_{\bf k}^c}{N!}\sum_{\sigma \in S_N}
\prod_{i<j}\left(1 -  \frac{i c\sign(x_j-x_i)}{k_{\sigma(j)}-k_{\sigma(i)}}\right)e^{i\sum_{j=1}^N x_j k_{\sigma(j)}}\;,
\\
&C_{\bf k}^c=\left(\prod_{i<j}\frac{(k_i-k_j)^2}{(k_{i}-k_{j})^2+c^2}\right)^{1/2}\;.
\end{align}
where the $|\Psi_{\bf k}^c\rangle$ denote the eigenstates of $\hat {\cal H}_N(c)$. 
The calculation is similar to the one in refs.~\cite{CDR10,dotsenko}
except that we
retain here for $c>0$ only the particle states.
Consider the imaginary time propagator at coinciding endpoints 
\be
Z_N({\bf 0},t|{\bf 0};c)
=\langle {\bf 0}|e^{-\hat{\cal H}_N(c) t}|{\bf 0} \rangle=  \int_{\mathbb{R}^N} \frac{\rmd {\bf k}}{(2\pi)^N}\langle {\bf 0}|\Psi^c_{\bf k}\rangle\langle \Psi^c_{\bf k}|{\bf 0}\rangle e^{-E_{\bf k} t}\;.
\ee
We now use that $\langle {\bf 0}|\Psi^c_{\bf k}\rangle=C_{\bf k}^c$ and we arrive at the following
 Laplace transform for the PDF of the coincidence time for $N$ Brownian bridges
\be\label{MGF_BB}
\moy{e^{-c {\cal T}_N(t)}}_{{\rm BB}}=\left(4 \pi t \right)^{\frac{N}{2}}\int_{\mathbb{R}^N} \frac{\rmd  {\bf k}}{(2 \pi)^N} e^{- {\bf k}^2 t} \prod_{i<j}\frac{(k_i-k_j)^2}{(k_i-k_j)^2+c^2}\;.
\ee
where we have divided by the result at $c=0$ which coincides with the free propagator $Z_N({\bf 0},t|{\bf 0};0)={(4\pi t)^{-\frac{N}{2}}}$. \\

We note that there is an alternative formula for the moments of the directed polymer problem, which was derived using Macdonald processes \cite{BorodinMacdo,G18}. Using this formula we can write the MGF of $N$ Brownian bridges equivalently as
\be\label{C_MGF}
\moy{e^{-c {\cal T}_N(t)}}_{\rm BB}=\left(4 \pi t \right)^{\frac{N}{2}} \int_{i\mathbb{R}^N}\frac{\rmd {\bf z}}{(2 i \pi)^N} e^{t{\bf z}^2} \prod_{i<j}\frac{z_i-z_j}{z_i-z_j+c}\;.
\ee
Note that this formula is valid only for $c>0$, otherwise the contours are different. Using the symmetrization identity
\be
\sum_{\sigma \in S_n} \prod_{i>j} \frac{z_{\sigma(i)} - z_{\sigma(j)} + c}{z_{\sigma(i)} - z_{\sigma(j)} }= n!
\ee 
one can show that this formula is in agreement with \eqref{MGF_BB}. 

\subsection{Probability Distribution Function of the coincidence time}\label{BB_PDF}

In this section we obtain from Eq. \eqref{MGF_BB} an expression for the PDF of the coincidence time for the
$N$ Brownian bridges. First, we use the Cauchy identity
\be
\prod_{i<j}\frac{(k_i-k_j)^2}{(k_i-k_j)^2+c^2}=\det_{1\leqslant \ell,m\leqslant N}\left(\frac{c}{c+i(k_\ell-k_m)}\right)= \sum_{\sigma \in S_N} \sign(\sigma) \prod_{\ell=1}^{N} \frac{c}{c+i(k_\ell-k_{\sigma(\ell)})}\;.
\ee
We then introduce the auxiliary variables ${\bf r}=(r_1,\cdots,r_N)$ such that
\begin{align}
\prod_{i<j}\frac{(k_i-k_j)^2}{(k_i-k_j)^2+c^2}&=c^N \sum_{\sigma \in S_N} \sign(\sigma)\int_{\mathbb{R}_+^{N}} \rmd  {\bf r}\; e^{-c\sum_\ell r_\ell} \prod_{\ell=1}^{N}e^{-ir_\ell(k_\ell-k_{\sigma(\ell)})}\\
&=c^N \sum_{\sigma \in S_N} \sign(\sigma)\int_{\mathbb{R}_+^{N}} \rmd  {\bf r}\; e^{-c\sum_\ell r_\ell} \prod_{\ell=1}^{N}e^{-ik_\ell(r_\ell-r_{\sigma(\ell)})}\;.
\end{align}
Replacing in Eq. \eqref{MGF_BB}, we are now able to compute the integrals over ${\bf k}$, yielding
\begin{align}
\moy{e^{-c {\cal T}_N(t)}}_{{\rm BB}}&=c^N \int_{\mathbb{R}_+^{N}} \rmd  {\bf r}\; e^{-c\sum_\ell r_\ell} \sum_{\sigma \in S_N} \sign(\sigma) \prod_{\ell=1}^{N} e^{-\frac{(r_\ell-r_{\sigma(l)})^2}{4t}}\\
&=c^N \int_{\mathbb{R}_+^{N}} \rmd  {\bf r}\; e^{-c\sum_\ell r_\ell} \det_{1\leqslant i,j\leqslant N}\left(e^{-\frac{(r_i-r_{j})^2}{4t}}\right)
\end{align}
Inverting the Laplace transform of this expression, we obtain the PDF $P_{N,{\rm BB}}(T)$ of the rescaled coincidence time ${\cal T}_N= {\cal T}_N(t=1)$ as
\begin{align}
P_{N,{\rm BB}}(T)&=\partial_T^N \int_{\mathbb{R}_+^{N}} \rmd {\bf r}\; \delta\left(T-\sum_\ell r_\ell\right) \det_{1\leqslant i,j\leqslant N}\left(e^{-\frac{(r_i-r_{j})^2}{4}}\right)\nn\\
&=\partial_T^{N+1} \int_{\mathbb{R}_+^{N}} \rmd {\bf r}\; \Theta\left(T-\sum_\ell r_\ell\right) \det_{1\leqslant i,j\leqslant N}\left(e^{-\frac{(r_i-r_{j})^2}{4}}\right)\label{PDF_P_BB}
\end{align}
where $\Theta(x)$ is the Heaviside step function. We start by analysing the case of $N=2,3$, where the PDF can be computed explicitly before considering the general behaviour for arbitrary values of $N$.  
This formula \eqref{PDF_P_BB} is the most compact general expression that we could find. We show in the next section that it can be used to derive explicitly the PDF for small values of $N=2,3$.

\subsection{Full distribution of the coincidence time for $N=2,3$ Brownian bridges}\label{BB_23}
\begin{figure}[h!]
\centering
\includegraphics[width=0.49\textwidth]{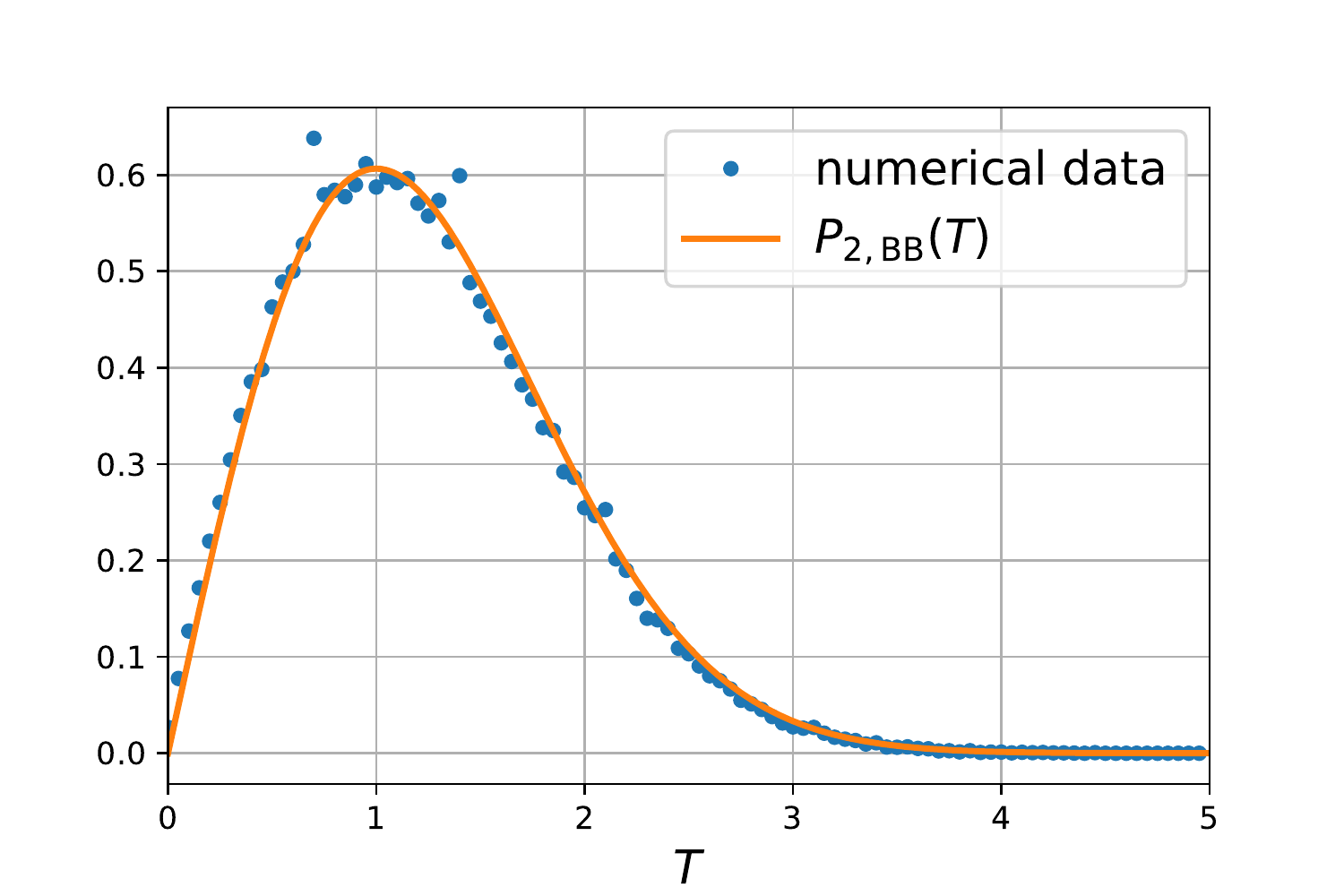}
\includegraphics[width=0.49\textwidth]{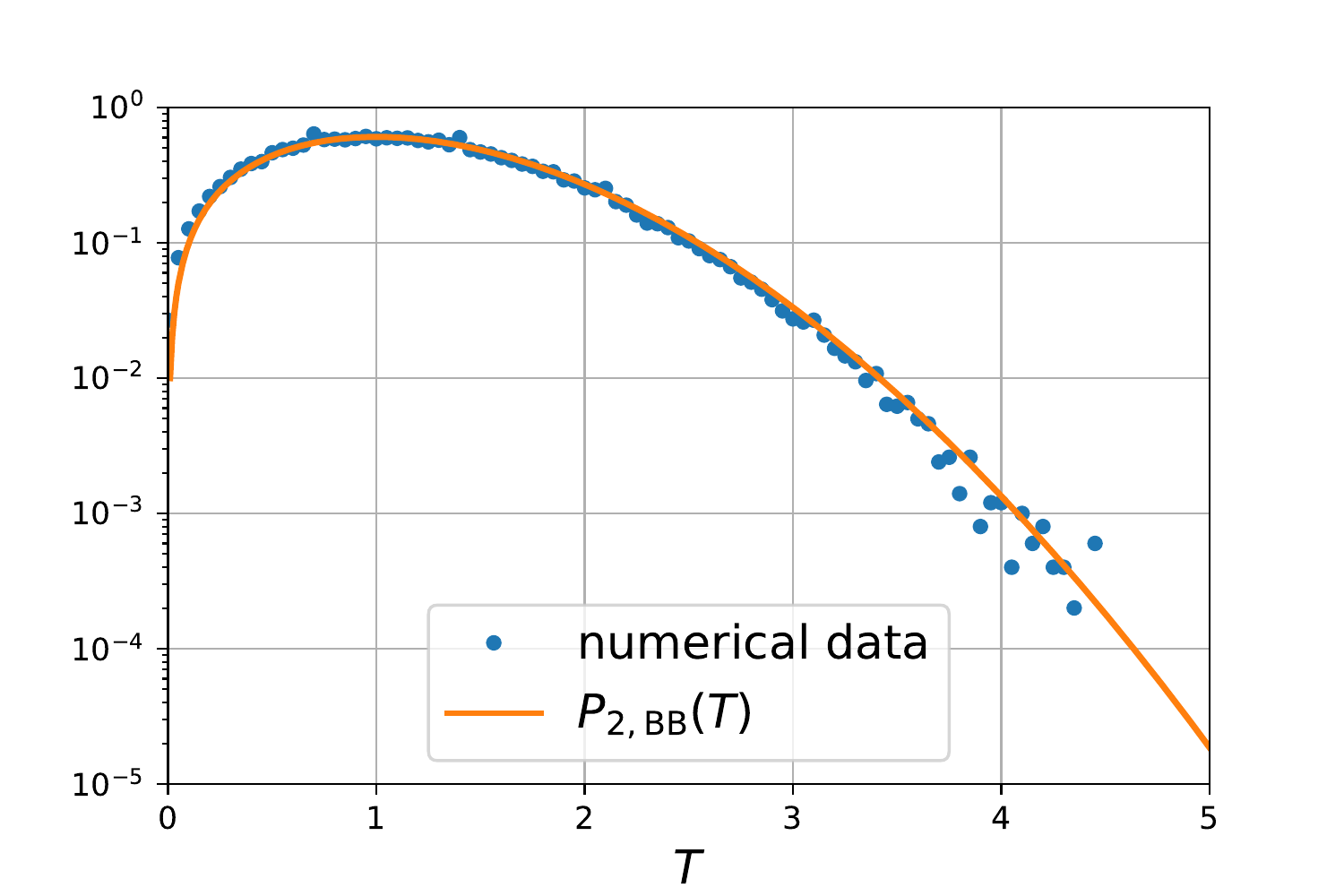}
\caption{Comparison between the PDF of the coincidence time obtained numerically for $N=2$ Brownian bridges on the time interval $\tau \in[0,1]$ and diffusion coefficient $D=1$ and the analytical result in Eq. \eqref{P_2_BB_ana} (some details on the simulations are provided in Appendix \ref{simul}). {\rm Left}: Linear scale, {\rm Right}: Logarithmic scale. }\label{fig_BB_2}
\end{figure}

~~~
\textbf{Distribution for $N=2$.} In the case of $N=2$, we will see how to extract the full PDF from Eq. \eqref{PDF_P_BB}.
Setting $N=2$, we may compute exactly the determinant in the integrand, yielding
\be\label{P_2_BB_nonex}
P_{2,{\rm BB}}(T)=\partial_T^{3} \int_{0}^{T}\rmd r_1\int_{0}^{T-r_1}\rmd r_2\left[1-e^{-\frac{(r_1-r_{2})^2}{2}}\right]=-\partial_T^{3}\int_{0}^{T}\rmd r_1\int_{0}^{T-r_1}\rmd r_2 e^{-\frac{(r_1-r_{2})^2}{2}}\;.
\ee
The first term in the integrand vanishes when deriving with respect to $T$. The PDF can then be derived in a few steps as 
\be\label{P_2_BB_ana}
P_{2,{\rm BB}}(T)=-\partial_T^{2} \int_{0}^{T}\rmd r e^{-\frac{(2r-T)^2}{2}}=
-\partial_T^{2} \int_{0}^{T}\rmd u 
e^{-\frac{u^2}{2}}
=
 ~T e^{-\frac{T^2}{2}}\;.
\ee
The asymptotic behaviours of this PDF are simple to extract as
\be\label{P_2_BB_asy}
P_{2,{\rm BB}}(T)=\begin{cases}
\displaystyle T+{\cal O}(T^3)&\;,\;\;T\to0\;,\\
&\\
T e^{-\frac{T^2}{2}} &\;,\;\;T\to \infty\;.
\end{cases}
\ee
In Fig. \ref{fig_BB_2}, we compare our analytical formula of Eq. \eqref{P_2_BB_ana} with the numerical simulations of Brownian bridges (see Appendix \ref{simul} for the details of the simulations), showing an excellent agreement. 

Note that in the case $N=2$, this result can be obtained directly from Eq. \eqref{P_joint_T_d} by setting the final distance between the diffusive particles $d=0$ and ensuring the normalisation of the probability
\be\label{P_2_BB_ana_2}
P_{2,{\rm BB}}(T)=\frac{P_{\rm joint}(T,d=0)}{\int_0^{\infty} P_{\rm joint}(T,d=0)\rmd T}=T e^{-\frac{T^2}{2}}\;.
\ee
As noted previously this method cannot be generalised to larger values of $N$, a contrario with our exact formula in Eq. \eqref{PDF_P_BB}.\\

\begin{figure}[h!]
\centering
\subfloat{%
\includegraphics[width=0.49\textwidth]{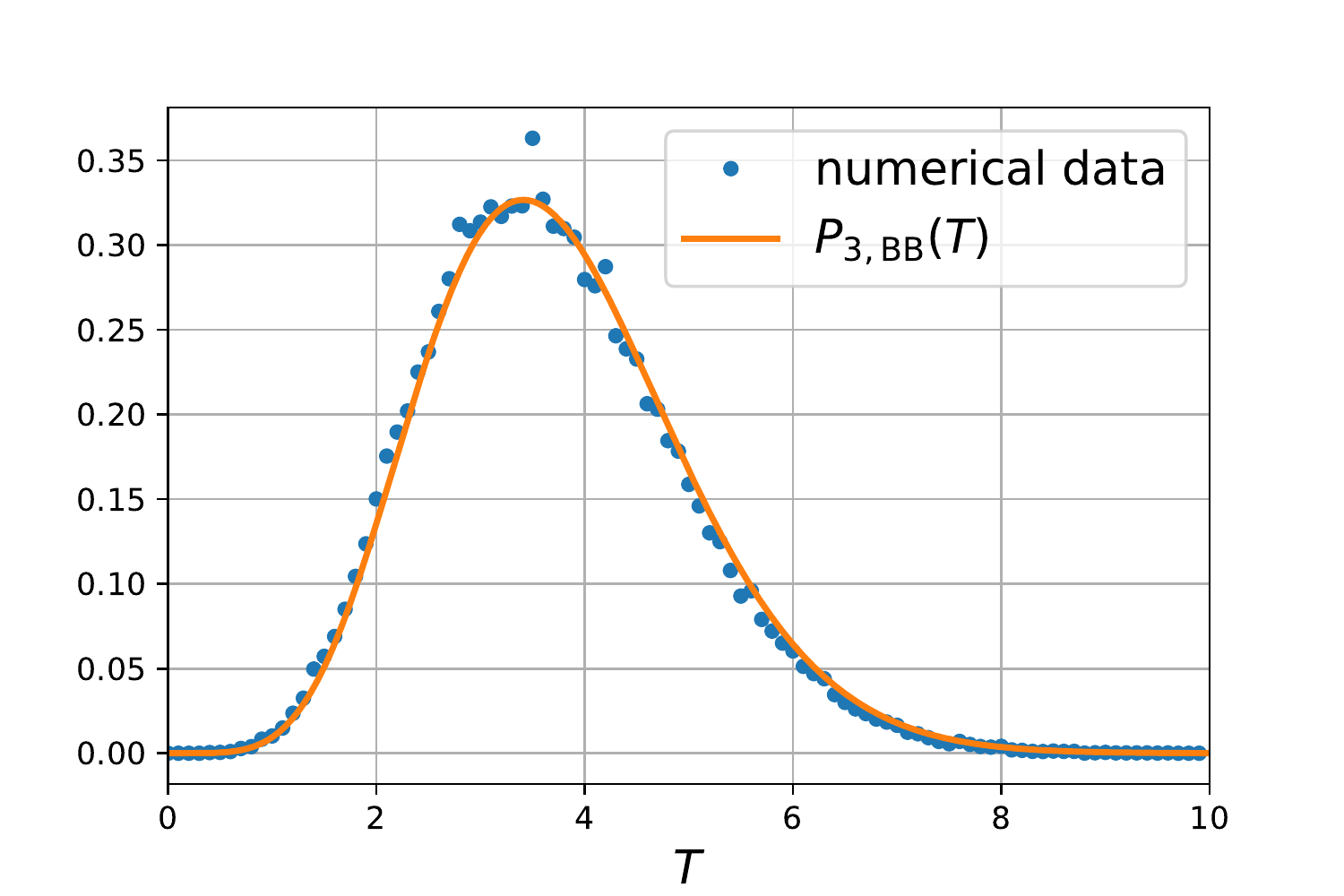}%
}
\subfloat{%
\includegraphics[width=0.49\textwidth]{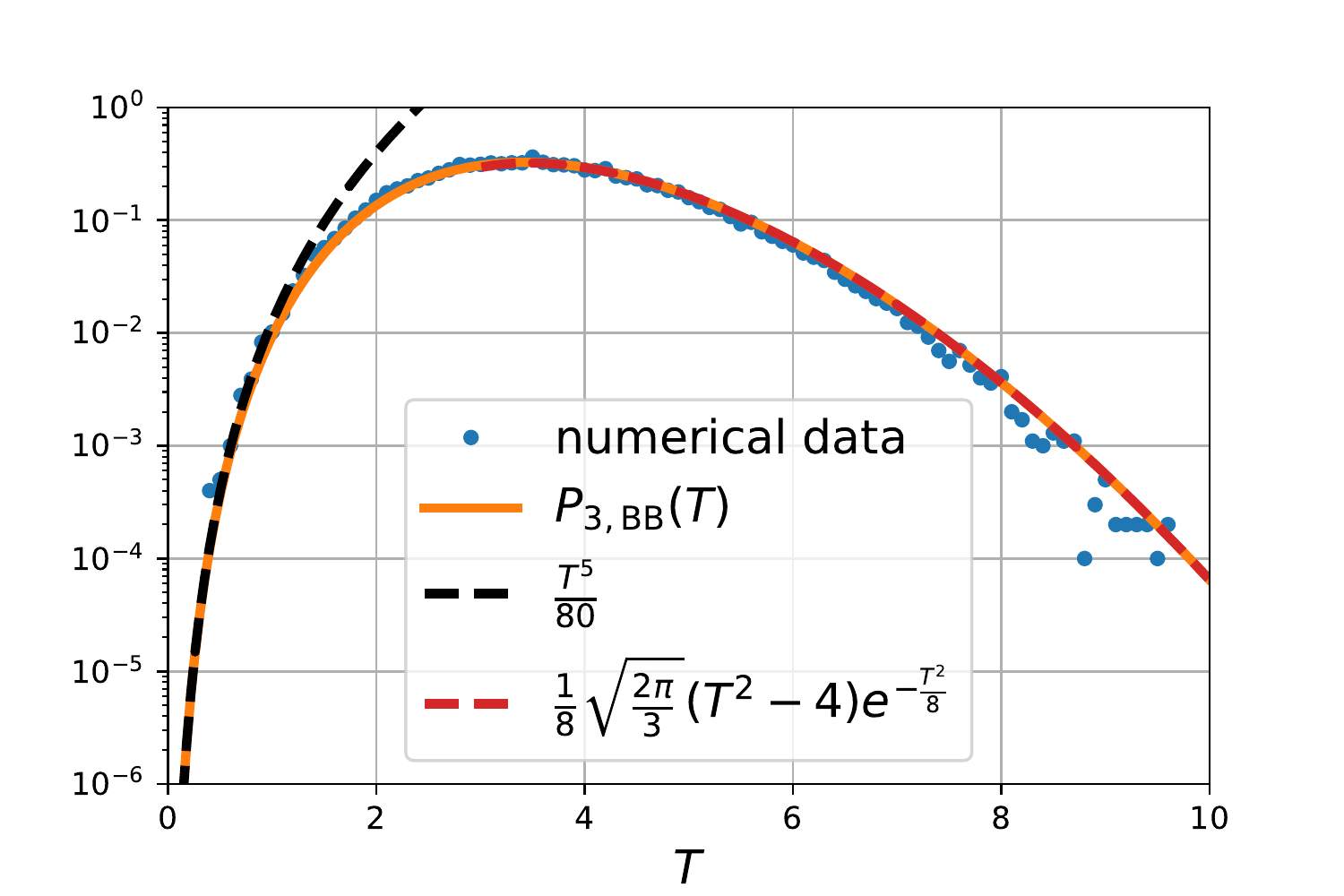}%
}
\caption{Comparison between the coincidence time obtained numerically for $N=3$ Brownian bridges on the time interval $\tau \in[0,1]$ and diffusion coefficient $D=1$ and the analytical result in Eq. \eqref{P_3_BB_ana}. {\rm Left}: Linear scale, {\rm Right}: Logarithmic scale. }\label{fig_BB_3}
\end{figure}

\textbf{Distribution for $N=3$.} Starting from $N=3$, the distribution cannot be obtained 
from a simple rescaling 
of the local time of a single Brownian motion,
 and we need to use our results in Eq. \eqref{PDF_P_BB} for the Brownian bridge. In the case of $N=3$, the determinant appearing in the integrand can still be computed quite easily, yielding
\be\label{P_3_BB_nonex}
P_{3,{\rm BB}}(T)=\partial_T^{4} \int_{0}^{T}\rmd r_1\int_{0}^{T-r_1}\rmd r_2\int_{0}^{T-r_1-r_2}\rmd r_3\left[1-
3e^{-\frac{(r_1-r_{2})^2}{2}}
+2\prod_{i=1}^3 e^{-\frac{(r_i-r_{i+1})^2}{4}}  
 \right]\;,
\ee
where we used the symmetry between the $r_i$'s. After some computations (See Appendix~\ref{PDF_3_ap} for details), we finally obtain the PDF
\be\label{P_3_BB_ana}
P_{3,{\rm BB}}(T)=\frac{T}{2} e^{-\frac{T^2}{2}}+\frac{1}{8}\sqrt{\frac{2\pi}{3}}e^{-\frac{T^2}{8}}(T^2-4)\erf\left(\sqrt{\frac{3}{8}}T\right)\;.
\ee
Its asymptotic behaviours read
\be
\label{eq:P3BB_asympt}
P_{3,{\rm BB}}(T)= \begin{cases}
\displaystyle\frac{T^5}{80}+\mathcal{O}(T^7)&\;,\;\;T\to 0\;,\\
&\\
\displaystyle\frac{1}{8} \sqrt{\frac{2\pi }{3}} 
   \left(T^2-4\right)e^{-\frac{T^2}{8}}+\mathcal{O}(e^{-\frac{T^2}{2}})&\;,\;\;T\to \infty\;.
\end{cases}
\ee
In Fig. \ref{fig_BB_3}, we compare the analytical formula of Eq. \eqref{P_3_BB_ana} with the numerical simulations of Brownian bridges (see Appendix \ref{simul} for the details of the simulations), showing an excellent agreement.

\subsection{Asymptotic behaviours of $P_{N,{\rm BB}}(T)$ for arbitrary $N$}\label{BB_asy}

We now come back to the case of $N$ Brownian bridges and analyse the tails of the PDF $P_{N,{\rm BB}}(T)$ of the rescaled coincidence time ${\cal T}_N={\cal T}_N(t)/\sqrt{t}$ respectively for $T\to 0$ and $T\to \infty$.

\subsubsection{Small $T$ limit of the Probability Distribution Function}

To obtain the small $T$ limit of $P_{N,{\rm BB}}(T)$, it is convenient to consider the $c\to +\infty$ limit of $\moy{e^{-c {\cal T}_N}}_{{\rm BB}}$, which is given in Eq. \eqref{MGF_BB}. The trajectories which contribute to a small coincidence time are repelled from each others. 
In the Lieb-Liniger picture, this is consistent with the repulsive case where the states are described as single particle states rather than bound state. 
%
%
Taking the large $c\to +\infty$ limit of Eq. \eqref{MGF_BB}, we obtain
\be\label{large_c}
\moy{e^{-c {\cal T}_N}}_{{\rm BB}}\approx \frac{1}{\pi^{\frac{N}{2}}}c^{-N(N-1)}\int_{\mathbb{R}^N} \rmd  {\bf k} 
e^{-{\bf k}^2} \prod_{i<j}(k_i-k_j)^2=2^{- \frac{N(N-1)}{2}} \frac{G(N+2)}{c^{N(N-1)}}\;,
\ee
where $G(n)=\prod_{k=1}^{n-2} k!$ is the Barnes $G$ function. 
We have used the
Mehta integral formula e.g. (1.5)-(1.6) in \cite{FoWa08}. Inverting the Laplace transform, we obtain
the small $T$ behavior
\be \label{small_TBB}
P_{N,{\rm BB}}(T) \underset{T\to 0}{=} 2^{- \frac{N(N-1)}{2}} \frac{G(N+2)}{\Gamma(N(N-1))}T^{N(N-1)-1}+\mathcal{O}(T^{N(N-1)+1}).
\ee
Note that setting $N=2,3$, using $G(4)=2$, $G(5)=12$ and $\Gamma(6)=120$, one recovers explicitly the result of the first line of Eqs. \eqref{P_2_BB_asy} and \eqref{eq:P3BB_asympt}.

\subsubsection{Large $T$ limit of the Probability Distribution Function}
\label{subsec:largeT_BB}

To study the large $T$ limit of $P_{N,{\rm BB}}(T)$, one needs
to investigate the $c\to -\infty$ limit of the expectation $\moy{e^{-c {\cal T}_N}}_{{\rm BB}}$ which is dominated
by large values of ${\cal T}_N$. The trajectories which contribute to a large coincidence time are attracted to each others. In the Lieb-Liniger picture, this corresponds to the attractive  case, where particles form bound states called strings. Upon increasing $\bar{c}=-c$, the potential becomes more and more attractive and the trajectories become dominated by the configuration where all the bosons are bounded into a single string.\\

Mathematically, for $c=-\bar c<0$ the expression of the moments of the directed polymer partition function is more involved as it involves a sum over string states \cite{CDR10,dotsenko,BorodinMacdo,sineG}.
Nonetheless, in the limit $\bar{c}\to +\infty$, the energy spectrum of the Lieb-Liniger model is dominated by its ground state which is a single string containing all particles with energy $E_0(N)=-\bar c^2 \frac{N(N^2-1)}{12}$. It follows from the contribution of the ground state
to the moment (see e.g. (65-66) in Supp. Mat. of \cite{KPZLateTime} replacing $t$ by $\bar c^2 t$ and 
multiplying by $c^N$) that
\begin{equation}
\moy{e^{-c {\cal T}_N}}_{{\rm BB}}= \frac{N!(4\pi)^{\frac{N-1}{2}}}{N^{3/2}} {\bar c}^{N-1}e^{{\bar c}^2 \frac{N(N^2-1)}{12}}[1+\mathcal{O}(e^{-\frac{1}{4}N(N-1)\bar{c}^2})]
\end{equation}
One can check that this behavior is consistent with the following $T\to \infty$ asymptotics 
\begin{align} 
P_{N,\rm{BB}}(T)&=\frac{N!2^{N-1}\pi^{\frac{N}{2}-1}}{N^{3/2}} \sqrt{\alpha_N}   (-\partial_T)^{N-1} e^{-\alpha_N T^2}+\mathcal{O}(e^{-\beta_N T^2})\label{PP} \\
&=\frac{N!2^{N-1}\pi^{\frac{N}{2}-1}}{N^{3/2}}\alpha_N^{\frac{N}{2}} H_{N-1}\left(\sqrt{\alpha_N}T\right)  e^{-\alpha_N T^2}+\mathcal{O}(e^{-\beta_N T^2})\;,
\end{align}
where $H_p(x)=e^{x^2}(-\partial_x)^p e^{-x^2}$ is the Hermite polynomial of degree $p$. The exponential factors are equal to 
\begin{equation}
\label{eq:largeT_expFactor}
\begin{cases}
\displaystyle \alpha_N=\frac{3}{N(N^2-1)}\\
\displaystyle \beta_N=\frac{3}{N^3-3N^2+2N}
\end{cases}
\end{equation}
This is checked by calculating the Laplace transform of \eqref{PP} using
a saddle point approximation (with a saddle point for  $T^*={\bar c}\alpha_N^{-1}>0$  for ${\bar c}=-c>0$).\\

For completeness, we provide this asymptotics explicitly for $N=2,3$.
\begin{itemize}
\item $N=2$, $P_{2,\rm{BB}}(T)=Te^{-\frac{T^2}{2}}$. 
This matches eq.~\eqref{P_2_BB_ana}.
\item $N=3$, $P_{3,\rm{BB}}(T)=\frac{1}{8} \sqrt{\frac{2\pi }{3}} e^{-\frac{T^2}{8}}
   \left(T^2-4\right)+\mathcal{O}(e^{-\frac{T^2}{2}})$. This matches eq.~\eqref{eq:P3BB_asympt}.
\end{itemize}

\subsection{Mean and variance of the coincidence time for $N$ Brownian bridges}\label{BB_mean_var}
In this section, we compute the two first cumulant of the Brownian bridge coincidence time ${\cal T}_N$ for arbitrary values of $N$.\\

\textbf{Mean value of ${\cal T}_{N,\rm BB}$} To compute the first moment ${\cal T}_{N,\rm BB}$, there are two alternative methods. We may obtain it by expanding Eq.~\eqref{C_MGF} for small $c$ or we may use the Brownian bridge propagator to compute it directly. In this section, we present both of these methods.
We start by considering Eq. \eqref{C_MGF}. 
We 
take a derivative of
 this equation with respect to $c$ to obtain 
\be\label{cum_C}
\moy{{\cal T}_N e^{-c {\cal T}_N}}_{\rm BB}=\left(4 \pi\right)^{\frac{N}{2}} \int_{i\mathbb{R}^N}\frac{\rmd {\bf z}}{(2 i \pi)^N} 
e^{{\bf z}^2}
 \prod_{i<j} \frac{(z_i-z_j)}{(z_i-z_j)+c}\sum_{i<j}\frac{1}{(z_i-z_j)+c}\;.
\ee
We introduce the auxiliary variables $r$ and take the limit $c\to 0^{+}$ to recast this integral as  
\be 
\moy{{\cal T}_N}_{\rm BB}=\left(4 \pi\right)^{\frac{N}{2}} \int_{i\mathbb{R}^N}\frac{\rmd {\bf z}}{(2 i \pi)^N} 
e^{{\bf z}^2}
\sum_{i<j}\int_0^{\infty}\rmd r e^{-(z_i-z_j+0^{+})r}\;.
\ee
where the $0^+$ term is explicitly written for convergence. We may then change variables from $z_\ell\to i k_\ell$ and compute the integrals for $\ell \neq i,j$, yielding
\begin{align} 
\moy{{\cal T}_N}_{\rm BB}&=4\pi\frac{N(N-1)}{2} \int_0^{\infty}\rmd r\int_{\mathbb{R}^2}
\frac{\rmd k_1 \rmd k_2 }{(2  \pi)^2} e^{-k_1^2-k_2^2}e^{-i(k_1-k_2)r-0^{+}r}\nn\\
&=\frac{N(N-1)}{2}\int_0^{\infty}\rmd r e^{-\frac{r^2}{2}}=\frac{N(N-1)}{2}\sqrt{\frac{\pi}{2}}\;.\label{mean_BB}
\end{align}
Alternatively, this computation can be done by introducing the Brownian bridge propagator
\be\label{BB_prop}
P_{\rm BB}(x,\tau|x_0,0,1)=\frac{e^{-\frac{(x-x_0)^2}{4D\tau(1-\tau)}}}{\sqrt{4\pi D\tau(1-\tau)}}\;.
\ee
where the symbol $P_{\rm BB}(x,\tau|x_0,0,1)$ denotes the propagator of the Brownian from $x_0$ at time $t=0$ to $x$ at time $\tau$ conditioned on being $x_0$ at time $t=1$. The mean value of the coincidence time of our process can simply be computed as (now taking $D=1$)
\begin{align}
\moy{{\cal T}_N}_{\rm BB}&=\int_0^{1}\rmd \tau \sum_{i\neq j} \moy{\delta(x_i(\tau)-x_j(\tau))}=\int_0
^{1}\rmd \tau \sum_{i\neq j} \int_{-\infty}^{\infty} P_{\rm BB}(x_i,\tau|x_0,0,1)^2 \rmd x_i \nn\\
&=N(N-1)\int_0^{1} \frac{\rmd \tau}{\sqrt{\tau(1-\tau)}} \int_{-\infty}^{\infty}\frac{\rmd x}{4\pi}e^{-\frac{x^2}{2}}=\frac{N(N-1)}{2}\sqrt{\frac{\pi}{2}}\;,
\end{align}
which coincides with the above result.

\textbf{Variance of ${\cal T}_{N,\rm BB}$} The second moment of the distribution can be obtained in a similar fashion either using the cumulant generating function in Eq. \eqref{C_MGF} or the Brownian bridge propagator in Eq. \eqref{BB_prop}.
Using the propagator we obtain that there are three different kinds of contributions
\begin{align}
&\moy{{\cal T}_N^2}=\int_0^{1}\rmd \tau_1 \int_0^{1}\rmd \tau_2 \sum_{i\neq j}\sum_{\ell\neq m} \moy{\delta(x_i(\tau_1)-x_j(\tau_1))\delta(x_m(\tau_2)-x_\ell(\tau_2))}\\
&=\frac{N!}{(N-4)!}\left[\int_0^{1}\rmd \tau \int_{-\infty}^{\infty} P_{\rm BB}(x,\tau|0,0)^2 \rmd x\right]^2\nn\\
&+8\frac{ N! }{(N-3)!}\int_0^{1}\rmd \tau_1 \int_{\tau_1}^{1}\rmd \tau_2  \int_{-\infty}^{\infty}\int_{-\infty}^{\infty} P_{\rm BB}(x,\tau_1|0,0,1)^2 \nn \\
&\hspace*{2cm}\times P(y-x,\tau_2-\tau_1|0,1)P_{\rm BB}(y,\tau_2|0,0,1) \rmd x\rmd y\nn\\
&+4\frac{ N!}{(N-2)!}\int_0^{1}\rmd \tau_1 \int_{\tau_1}^{1}\rmd \tau_2  \int_{-\infty}^{\infty}\int_{-\infty}^{\infty} P(x,\tau_1|0,0)^2 P(y-x,\tau_2-\tau_1|0,1)^2 \rmd x\rmd y\;,\nn
\end{align}
where $P(x,\tau|0,0)=(4\pi\tau)^{-1/2}e^{-x^2/(4\tau)}$ is the free Brownian propagator. The first term in this expansion gives a disconnected contribution $\frac{N!}{4(N-4)!}\moy{{\cal T}_2}^2$. After a careful computation, we obtain the variance for the Brownian bridge
\be
\var{{\cal T}_N}_{\rm BB}=\moy{\left[{\cal T}_N-\moy{{\cal T}_N}_{\rm BB}\right]^2}_{\rm BB}=\frac{N!}{(N-3)!}\left[\frac{8\pi}{9\sqrt{3}}-\frac{\pi}{2}\right]+\frac{N!}{(N-2)!}\left[1-\frac{\pi}{4}\right]\;.\label{var_BB}
\ee
This is the last result that we obtain for arbitrary $N$ Brownian bridges.


In the limit of a large number of walkers, $N \to + \infty$ limit we find 
\be 
\moy{{\cal T}_N}_{\rm BB} \simeq 
\frac{1}{2} \sqrt{\frac{\pi}{2}} N^2 \quad , \quad 
 \var{{\cal T}_N}_{\rm BB} \simeq 
\left(\frac{8 \pi }{9 \sqrt{3}}-\frac{\pi }{2}\right) N^3 
\ee
This scaling with $N$ 
is characteristic of the regime of typical fluctuations.
We will now explore another regime, the large deviation regime, dominated 
by rare fluctuations. 

\subsection{Coulomb gas method for $N\to \infty$}\label{BB_CG}

We now study the limit of a large number of walkers $N\to \infty$. We 
will study the moment generating function of the coincidence time ${\cal T}_N$,
in that limit for $c>0$. As we show below, there is a natural scaling regime, $c \sim \sqrt{N}$,
which allows to use a Coulomb gas approach. This regime corresponds to large positive $c$, 
hence to events 
where the coincidence time ${\cal T}_N$ is much smaller than its average $\langle {\cal T}_N \rangle$. Exponentiating the integrand  of eq.~\eqref{MGF_BB}, we get
\be\label{MGF_BB2}
\moy{e^{-c {\cal T}_N(t)}}_{{\rm BB}} = \pi^{-N/2}
\int_{\mathbb{R}^N} \rmd  {\bf k}  e^{- {\cal S}_N({\bf k}) }
\ee
where the action is given by
\be
{\cal S}_N({\bf k}) = \sum_i k_i^2 + \sum_{i<j} \log\left[ 1 + \frac{c^2}{(k_i-k_j)^2}\right]
\ee
For the two terms to be of the same order we must scale $k_i=\sqrt{N} p_i$ and $c = \sqrt{N}\tilde{c}$. Introducing the empirical density
$\rho(p) = \frac{1}{N} \sum_i \delta(p-p_i)$ we rewrite the action as
\be
{\cal S}_N({\bf k}) = N^2  S_{\tilde c}[\rho] + o(N^2) 
\ee
where we have defined the Coulomb gas energy functional
\be\label{S}
S_{\tilde c}[\rho]=\int_{\mathbb{R}} \rmd p \, p^2 \rho(p)+\frac{1}{2}\iint_{\mathbb{R}^2} \rmd p \rmd p' \rho(p)\rho(p')\log\left[1+\frac{{\tilde c}^2}{(p-p')^2}\right]\;.
\ee
We can evaluate the integral in \eqref{MGF_BB2} using a saddle point approximation and obtain
\be \label{largedev1} 
\moy{e^{-c {\cal T}_N(t)}}_{{\rm BB}}\propto \exp\left[-N^2 \Psi\left(\frac{c}{\sqrt{N}}\right)\right]\;,
\ee
where from its definition the rate function $\Psi(\tilde c)$ is defined on $\tilde c \in [0,+\infty[$ and must 
be positive, increasing $\Psi'(\tilde c) \geqslant 0$ and concave $\Psi''(\tilde c) \leqslant 0$, with $\Psi(0)=0$. 
It is determined  by minimizing the Coulomb gas energy
\be
\Psi(\tilde c)=\min_{\rho} \left[ S_{\tilde c}[\rho] - \mu\left(\int_\mathbb{R} \rho(p) \rmd p - 1\right) \right]
\ee
with respect to the density $\rho(p)$, where we have introduce a Lagrange
multiplier $\mu$ to enforce the normalization constraint $\int_\mathbb{R} \rmd p \rho(p)=1$. We denote $\rho^*({\tilde c},p)$ the minimizer of the action $S_{\tilde c}[\rho]$ under the normalisation constraint. Note that for $\tilde c=0$, there is no normalised density allowing to minimise $S[\rho]$. Computing the functional derivative, we obtain an integral equation for $\rho^*({\tilde c},p)$, 
valid for $p$ in the support of $\rho^*(\tilde c;p')$
\be\label{min_rho}
p^2+\int_\mathbb{R} \rmd p' \rho^*(\tilde c;p')\log\left[1+\frac{{\tilde c}^2}{(p-p')^2}\right]= \mu({\tilde c}) \;.
\ee
Multiplying this equation with $\rho^*(\tilde c;p)$ and integrating with respect to $p$, we obtain
\be
\iint_{\mathbb{R}^2} \rmd p \rmd p' \rho^*(\tilde c;p)\rho^*(\tilde c;p')\log\left[1+\frac{{\tilde c}^2}{(p-p')^2}\right]=\mu({\tilde c})-\int_{\mathbb{R}} \rmd p \rho^*(\tilde c;p)p^2 \;.
\ee
Replacing in Eq. \eqref{S}, this yields a simpler expression, which can be interpreted as a virial theorem, for the rate function $\Psi(\tilde c)$ for general value of $\tilde c$
\be \label{virial} 
\Psi(\tilde c)=\int_\mathbb{R} \rmd p \frac{p^2}{2}\rho^*(\tilde c;p)+\frac{\mu({\tilde c})}{2}\;.
\ee
Note also that 
taking a derivative of  Eq. \eqref{min_rho} with respect to $p$, we obtain an equation that does not depend on the Lagrange multiplier $\mu$ and reads, 
for any $p$ in the support of $\rho^*(\tilde c;p')$
\be
p=\dashint_\mathbb{R} \frac{\rho^*(\tilde c;p')\rmd p'}{p-p'}\frac{{\tilde c}^2}{{\tilde c}^2+(p-p')^2}\;.
\ee
Solving this integral equation for arbitrary values of $\tilde{c}$ is quite
non-trivial and left for future studies. Here we will only 
 solve this equation perturbatively in the regime $\tilde c\to \infty$. The density $\rho^*(\tilde c;p)$ can be expressed as a perturbative series of ${\tilde c}^{-2}$
\be
\rho^*(\tilde c;p)=\rho_{\rm sc}(p)+\frac{1}{{\tilde c}^2}\delta\rho_1(p)+\mathcal{O}(\frac{1}{\tilde c^{4}}),
\ee
where the leading order of the density $\rho_{\rm sc}(p)$ does not depend on $\tilde c$. It is solution of the integral equation
\be\label{eq_sc}
\dashint_\mathbb{R} \frac{\rho_{\rm sc}(p')\rmd p'}{p-p'}=p\;.
\ee
The general solution of the equation $\dashint_a^b \frac{\rho(p')\rmd p'}{p-p'}=g(p)$ 
(i.e. assuming
a single interval support)
can be obtained by the Tricomi formula \cite{Tri} 
\be\label{tri_for}
\rho(p)=\frac{1}{\pi\sqrt{p-a}\sqrt{b-p}}\left[\int_{a}^b \rho(p')\rmd p'-\dashint_a^b \frac{\rmd p'}{\pi} \frac{\sqrt{p'-a}\sqrt{b-p'}}{p-p'}g(p')\right]\;.
\ee
We expect in this case that $a=-b$ and that the density $\rho_{\rm sc}(p)$ is normalised. Inserting these two conditions, we obtain
\be
\rho(p)=\frac{2+a^2-2p^2}{2\pi\sqrt{a^2-p^2}}\;,\;\;-\sqrt{a}\leqslant p\leqslant \sqrt{a}\;.
\ee
Imposing that this density vanishes at the edges in $\pm a$, we obtain $a=\sqrt{2}$. The solution of Eq. \eqref{eq_sc} is then the Wigner semi-circle law
\be
\rho_{\rm sc}(p)=\frac{\sqrt{2-p^2}}{\pi}\;,\;\;-\sqrt{2}\leqslant p\leqslant \sqrt{2}\;.
\ee
Computing the next order term, we obtain for $p \in [-\sqrt{2},\sqrt{2}]$
\be
 \dashint_{-\sqrt{2}}^{\sqrt{2}} \frac{\delta\rho_{1}(p')\rmd p'}{p-p'}= \int_{-\sqrt{2}}^{\sqrt{2}} \rho_{\rm sc}(p')(p-p')\rmd p'=p\;. 
\ee
The solution at first order of perturbation must have a vanishing integral $\int_{-\sqrt{2}}^{\sqrt{2}}\delta\rho_{1}(p)\rmd p$ for the density to be normalised. Using again the Tricomi formula in Eq. \eqref{tri_for}, this yields
\be
 \delta\rho_{1}(p)=\frac{1-p^2}{\pi \sqrt{2-p^2}}\;,\;\;-\sqrt{2}\leqslant p\leqslant \sqrt{2}\;. 
\ee
We may then compute the rate function at order ${\tilde c}^{-2}$. 
Evaluating the Lagrange multiplier from Eq. 
 \eqref{min_rho} at, e.g. $p=0$, we obtain
\begin{equation}
\begin{split}
\mu(\tilde c)&=2\log \tilde c-2\int_\mathbb{R} \rmd p \rho_{\rm sc}(p)\log|p|+\frac{1}{{\tilde c}^2}\left[\int_\mathbb{R} \rmd p \rho_{\rm sc}(p) p^2-2\int_\mathbb{R} \rmd p \delta\rho_{1}(p)\log|p|\right]+\mathcal{O}({\tilde c}^{-4})\\
&=2\log \tilde c+1+\log 2+ \frac{1}{{\tilde c}^2} (\frac{1}{2} + 1 ) +\mathcal{O}({\tilde c}^{-4}).
\end{split}
\end{equation}
where we used the large $\tilde c$ expansion
\be
\log\left[1+\frac{{\tilde c}^2}{(p-p')^2}\right] = 2\log \tilde c - 2 \log|p-p'| + \frac{1}{\tilde c^2} (p-p')^2 
+ O(\frac{1}{\tilde c^4})
\ee

\begin{figure}[h!]
\centering
\includegraphics[width=0.8\textwidth]{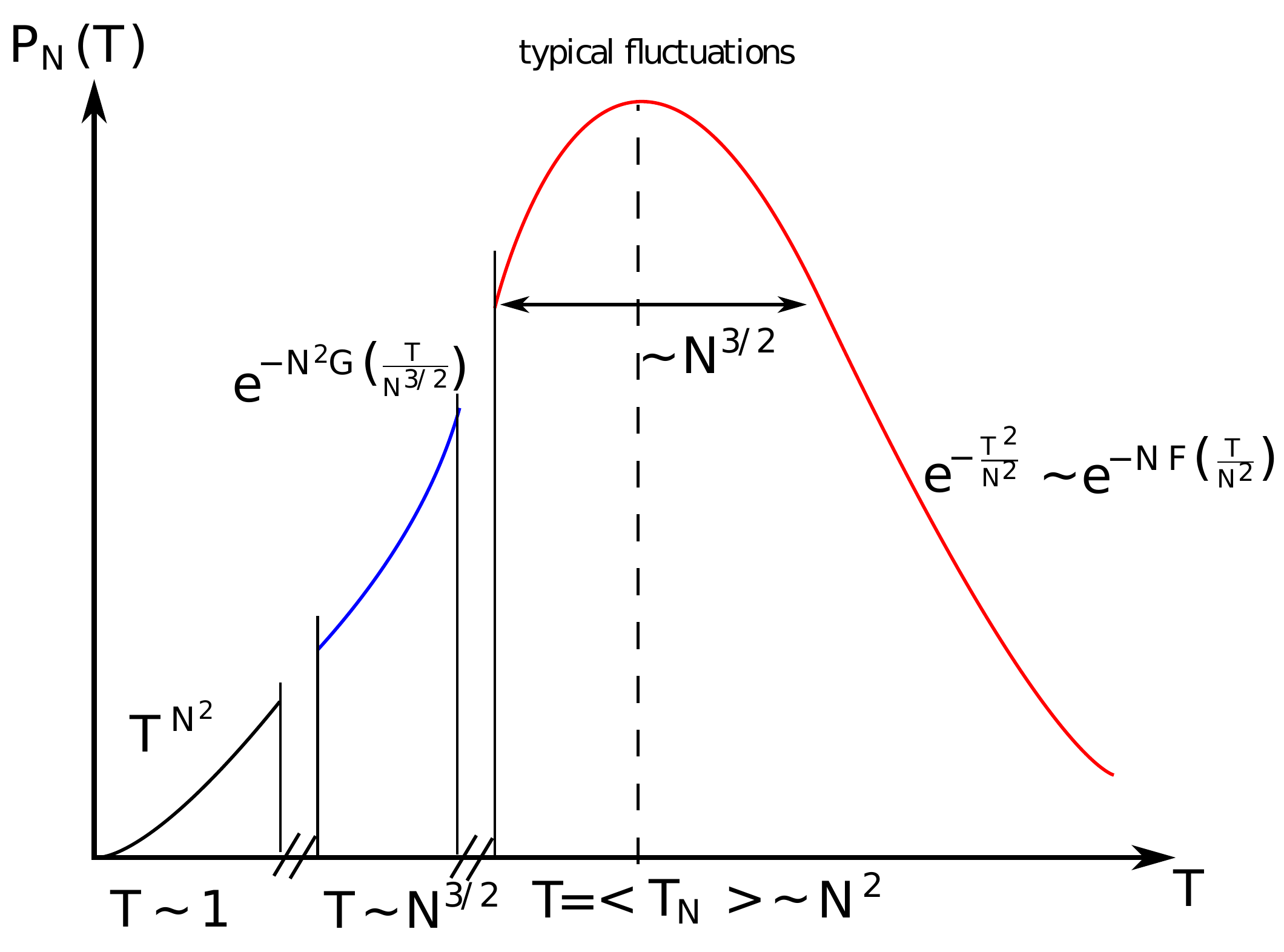}
\caption{
Schematic representation of the different regimes for the PDF of the coincidence
time $T$, $P_N(T)$, in the limit of large $N$. 
The average $T$ is $\langle {\cal T}_N \rangle \sim N^2$, and the typical fluctuations live in a window
$|T- \moy{{\cal T}_N}|\sim N^{3/2}$ around it. The large deviation regime studied in the text via the Coulomb gas is represented in blue and correspond to
$T\sim N^{3/2}\ll \moy{{\cal T}_N}$ and has rate $N^2$.   We have shown that its left part, which corresponds to
$T \ll N^{3/2}$ (and $\tilde c \gg 1$ for the corresponding MGF) matches the regime $T \sim 1$ which is also indicated. 
 The matching of its right part, i.e. $N^{3/2} \ll T \ll N^2$  (and $\tilde c \ll 1$ for the corresponding MGF),
towards the typical regime remains to
be studied. 
We note that the right tail of the regime of typical fluctuations is consistent with the existence of a right large deviation regime for $|T- \moy{{\cal T}_N}|\sim N^{2}$ of rapidity $N$, which remains to be studied.}\label{Figs}
\end{figure}

Gathering the different terms, we obtain from \eqref{virial} the large $\tilde c$ behavior

\be \label{resPsi} 
\Psi(\tilde c)= \log \tilde c+\frac{3}{4}+\frac{1}{2}\log 2+ \frac{1}{2 \tilde c^2} 
+ O(\frac{1}{\tilde c^4})
\ee
which is consistent, to leading order with $\Psi(\tilde c)$ being increasing 
and concave.  This behaviour, obtained here in the regime $N,c \to +\infty$ with large  $\tilde c=c/\sqrt{N}$,
can be compared with the large $c$ expansion at fixed $N$ that we obtained in Eq. \eqref{large_c},
\begin{align}
-\frac{1}{N^2}\log \moy{e^{-c {\cal T}_N}}_{{\rm BB}}
& \simeq_{c \to +\infty} -\frac{\log G(N+2)}{N^2}+\frac{N(N-1)}{N^2} (\log c+\frac{1}{2}\log 2)  \\
& \simeq_{N \to +\infty} \log \left[\frac{c}{\sqrt{N}}\right]+\frac{3}{4}+\frac{1}{2}\log 2
\end{align}
where we used the asymptotic behaviour of the Barnes function 
$\log G(z+2)= (z^2/2)\log z-3z^2/4+\mathcal{O}(z)$. This estimate
coincides precisely with the first three leading terms at large $\tilde c$ in \eqref{resPsi}. 
The matching of these two regimes is illustrated in the Figure \ref{Figs}.\\

If we assume that the above variational problem has a well behaved solution for any $\tilde c$,
then the MGF has the large deviation form Eq. \eqref{largedev1}. That is then 
consistent with the PDF of $T$ having the following large deviation tail in the regime 
$T \sim N^{3/2}$
\be
P_{N,{\rm BB}}(T) \sim e^{- N^2 G(\tilde T) } \quad , \quad \tilde T=\frac{T}{N^{3/2}} \;.
\ee
Indeed substituting this form in the definition of the Laplace transform, the integral
is dominated at large $N$ by the maximum of its integrand
\be
\int_0^{+\infty} \rmd T P_{N,{\rm BB}}(T) \, e^{-c T} \sim 
\int_0^{+\infty} \rmd T \, e^{- N^2 ( G(\tilde T) + \tilde c \tilde T ) }
\sim e^{- N^2 \min_{\tilde T \geqslant 0} [ G(\tilde T) + \tilde c \tilde T ] }
\ee 
Hence the 
rate function $G(\tilde T)$ is then determined via a Legendre transform
\be
\Psi(\tilde c)  = \min_{\tilde T  \geqslant 0} \, [ G(\tilde T) + \tilde c ~ \tilde T ]
\quad , \quad G(\tilde T) = \max_{\tilde c \geqslant 0}[ \Psi(\tilde c) - \tilde c ~ \tilde T]
\ee 
Denoting the saddle points $\tilde T^*({\tilde c})$ and $\tilde c^*({\tilde T})$ we 
have $\tilde T^*({\tilde c}) \simeq \frac{1}{\tilde c}- \frac{1}{\tilde c^3}$
at large $\tilde c$ and $\tilde c^*({\tilde T}) \simeq \frac{1}{\tilde T} - \tilde T$ at small
$\tilde T$. We obtain the small $\tilde T$ expansion
\be
G(\tilde T) = \log(\frac{1}{\tilde T}) - \frac{1}{4}+\frac{1}{2}\log 2 + \frac{\tilde T^2}{2} + o(\tilde T^2)
\ee 
Again, the first three terms match exactly the large $N$ limit of 
\eqref{small_TBB}. \\

The large deviation regime studied above thus corresponds to events of probability $- \log P = \mathcal{O}(N^2)$
in the double limit $N,T \to +\infty$ with $T \sim N^{3/2}$, \textit{i.e.} to coincidence times $T$ much smaller than the average (and typical) one $T=\moy{{\cal T}_N}_{\rm BB}$. It equivalently corresponds to configurations of Brownian paths which strongly repel each others, \textit{i.e.} in the Laplace variable this is the regime $c \sim \sqrt{N} >0$. This is illustrated in Figure \ref{Figs}.

It is then clear from the figure that the result \eqref{small_TBB} for the PDF, which corresponds to small  $T=\mathcal{O}(1)$ and fixed $N$, should match for large $N$ the large $\tilde c$ limit. One cannot exclude however
that there exists intermediate deviation regimes, \textit{e.g.} for $T \sim \moy{{\cal T}_N}_{\rm BB} \sim N^2$. 

One notes for instance  that the upper tail at fixed $N$ and large $T$ behaves as $P_N(T) \sim e^{- T^2/N^3}$ 
(see formula \eqref{PP}). Inserting $T \sim \moy{{\cal T}_N}_{\rm BB}$ we see that 
this factor in the upper tail can be of order
$e^{- \mathcal{O}(N)}$ at most. Studying this regime at large $N$ would require to extend the above Coulomb gas on the $c<0$ side, which is left for future studies.

\section{Coincidence time for Brownian motions and KPZ equation with flat initial condition}\label{sec:BM}

Let us now turn to the case of the Brownian motion, where the final points are not fixed
(i.e. they are integrated upon). This is connected to the directed polymer with one free endpoint,
equivalently with the Kardar-Parisi-Zhang equation with a flat initial condition. A solution for the
latter was given in \cite{we-flat,we-flatlong}. We will use this solution here both
for $c>0$ and $c<0$. Since it was obtained by first calculating the moments
for the half-flat initial condition we start by the latter, which corresponds
to $N$ Brownians all starting at $0$ and ending on a given half line. 

\subsection{Bethe ansatz solution of the Lieb-Liniger model for half-flat initial conditions}

Consider the partition sum of the directed polymer with half-flat initial condition
\be
\mathcal Z_w(x,t)= \int_{-\infty}^0 \rmd y e^{w y} {\cal Z}(x ,y, t) 
\ee 
From Eq (52) in \cite{PLDCrossoverDropFlat}, or Eq. (88) in \cite{we-flatlong}, we have
the following: restoring the factors of $\bar c$ (by a change of units), with $c=-\bar c$,
restricting to $c\geqslant 0$ and retaining only the single particle states (as they are the only states of the repulsive Lieb-Liniger model), we obtain
\bea
\label{eq:MGF_BM}
\fl \overline{\mathcal Z_w(x,t)^N} =  \prod_{j=1}^N \int_\mathbb{R} \frac{\rmd k_j}{2 \pi}  \frac{e^{- k_j^2 t - i x k_j}}{i k_j + w}  
\prod_{1 \leqslant i<j \leqslant N} \frac{(k_i-k_j)^2}{(k_i-k_j)^2 + c^2}  \frac{ i k_i + i k_j + 2 w + c}{i k_i + i k_j + 2 w} 
\eea 
From this we obtain the MGF for the coincidence time of $N$ Brownian walkers which start at $x$ and
reach the half line $\left]-\infty,0\right]$, as
\bea
\moy{e^{-c {\cal T}_N}}_{{\rm half-BM}} = \frac{\overline{\mathcal Z_w(x,t)^N}}{\mathcal Z^0_w(x,t)^N}
\quad , \quad \mathcal Z^0_w(x,t) = \int_\mathbb{R} \frac{\rmd k}{2 \pi}  \frac{e^{- k^2 t - i x k}}{i k + w} 
\eea

It is written here in presence of an extra weight $e^{w x}$, which can then be considered in
the limit $w=0^+$.

\subsection{Flat initial conditions as a limit of the half-flat}
The moments of the partition sum of the directed polymer with one free endpoint
i.e. $\mathcal Z_{\rm flat}(t) = \int_\mathbb{R} \rmd y {\cal Z} (x,y,t)$, associated to the Kardar-Parisi-Zhang equation with 
flat initial condition, can be obtained from the ones for
the half-flat initial condition in the double limit $x \to - \infty$ and $w \to 0^+$, as performed in 
\cite{we-flatlong}. In that limit only paired strings and strings with zero momenta remain:
this provides the MGF for the coincidence time of unconstrained Brownians walkers
all starting at $0$.

\subsubsection{Moment generating function and probability distribution function for $N=2,3$}

We start with the lowest moments, from eqs.~(52) and (56) in \cite{we-flatlong},
and consider $c>0$, in which case we keep only the contribution of 
single particle states. The first moment is simply $\overline{ \mathcal Z_{\rm flat}(t) }=1$ for all times $t$. We consequently focus on the moments at time $t=1$.\\

\begin{figure}[h!]
\centering
\includegraphics[width=0.49\textwidth]{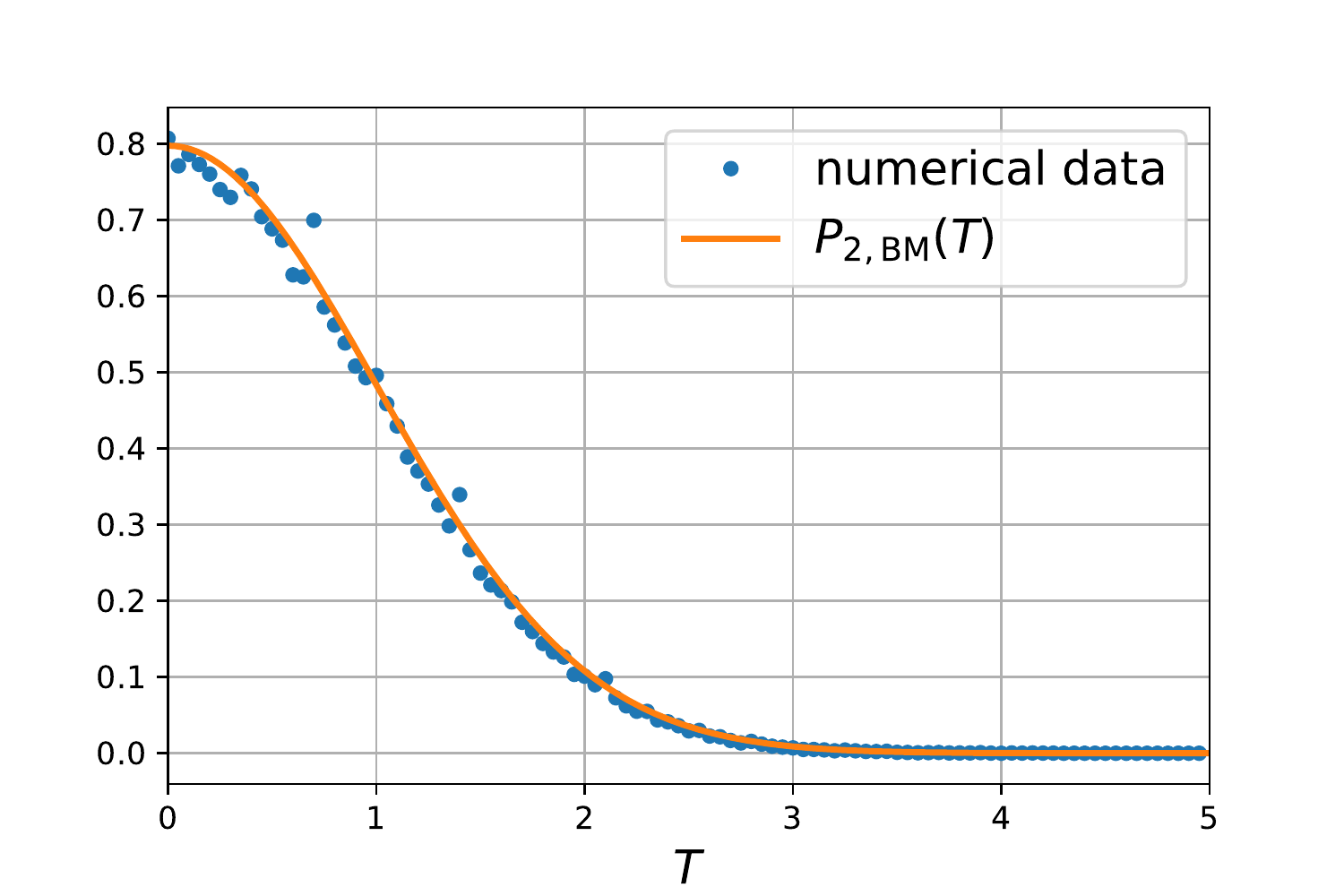}
\includegraphics[width=0.49\textwidth]{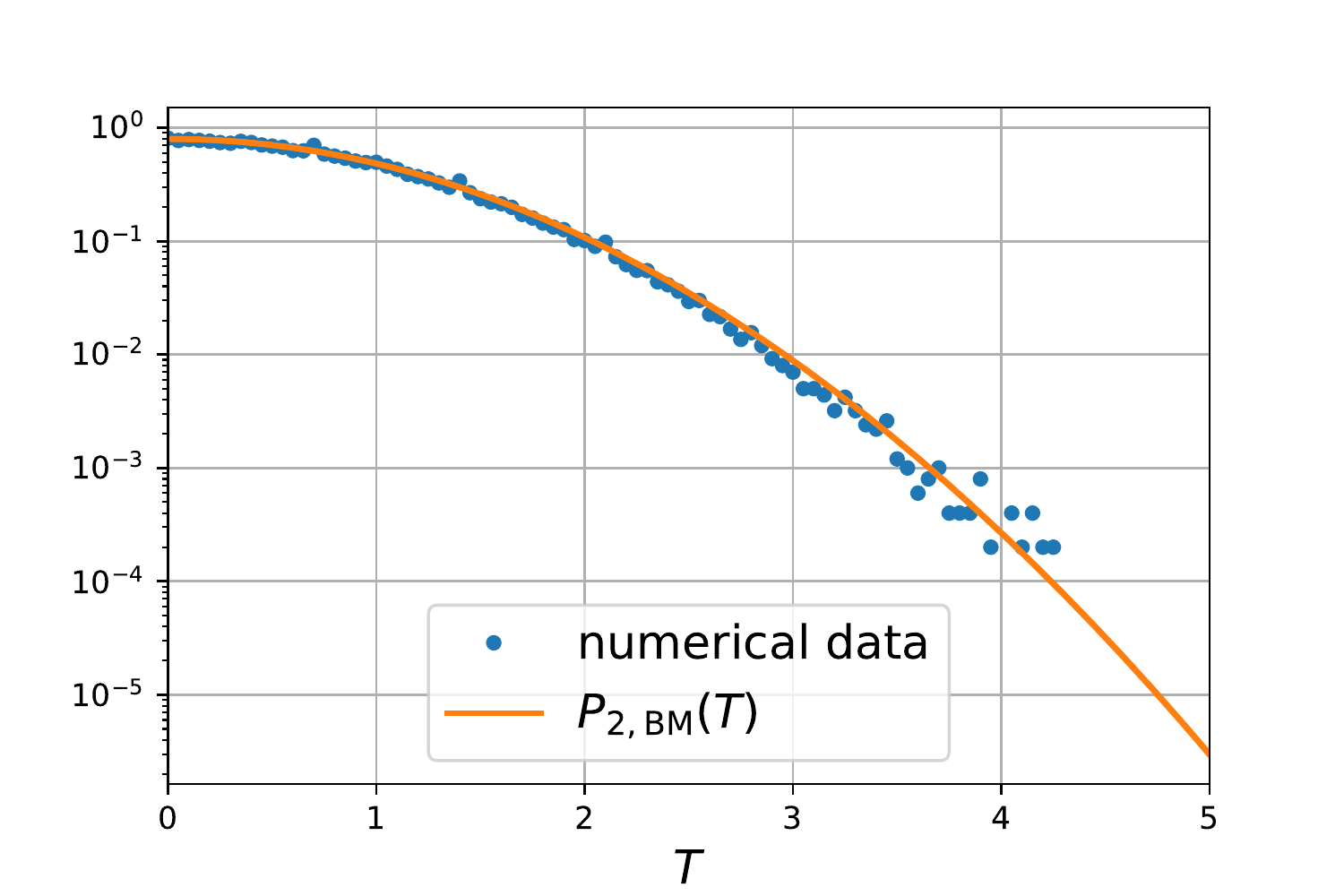}
\caption{Comparison between the coincidence time obtained numerically for $N=2$ Brownian motions on the time interval $\tau \in[0,1]$ and diffusion coefficient $D=1$ and the analytical result in Eq. \eqref{P_2_BM_ana} (some details on the simulations are provided in Appendix \ref{simul}). {\rm Left}: Linear scale, {\rm Right}: Logarithmic scale.}\label{P_2_BM}
\end{figure}

$\bm{N=2}$ From the second moment we obtain the result for two walkers
\be
\moy{e^{-c {\cal T}_2}}_{{\rm BM}}= \overline{\mathcal Z_{\rm flat}(1)^2 } = 4 c \int_\mathbb{R} \frac{\rmd k}{2 \pi} \frac{e^{-2 k^2}}{4 k^2 + c^2} = 
e^{\frac{c^2}{2}  } (1- \erf( \frac{c}{\sqrt{2}} ))
\ee 
The inverse Laplace transform then yields 
\begin{equation}
\begin{split}
P_{2,{\rm BM}}(T)=\mathcal{L}^{-1} \moy{e^{-c {\cal T}_2}}_{{\rm BM}}  &=  \int_\mathbb{R} \frac{\rmd k}{2 \pi} e^{-2 k^2} 4 \cos(2 k T) = \sqrt{\frac{2}{\pi}} e^{- \frac{T^2}{2}} \label{P_2_BM_ana}
\end{split}
\end{equation}
This result compares very well with the numerical simulations in Fig. \ref{P_2_BM}.

\begin{figure}[h!]
\centering
\subfloat{%
\includegraphics[width=0.49\textwidth]{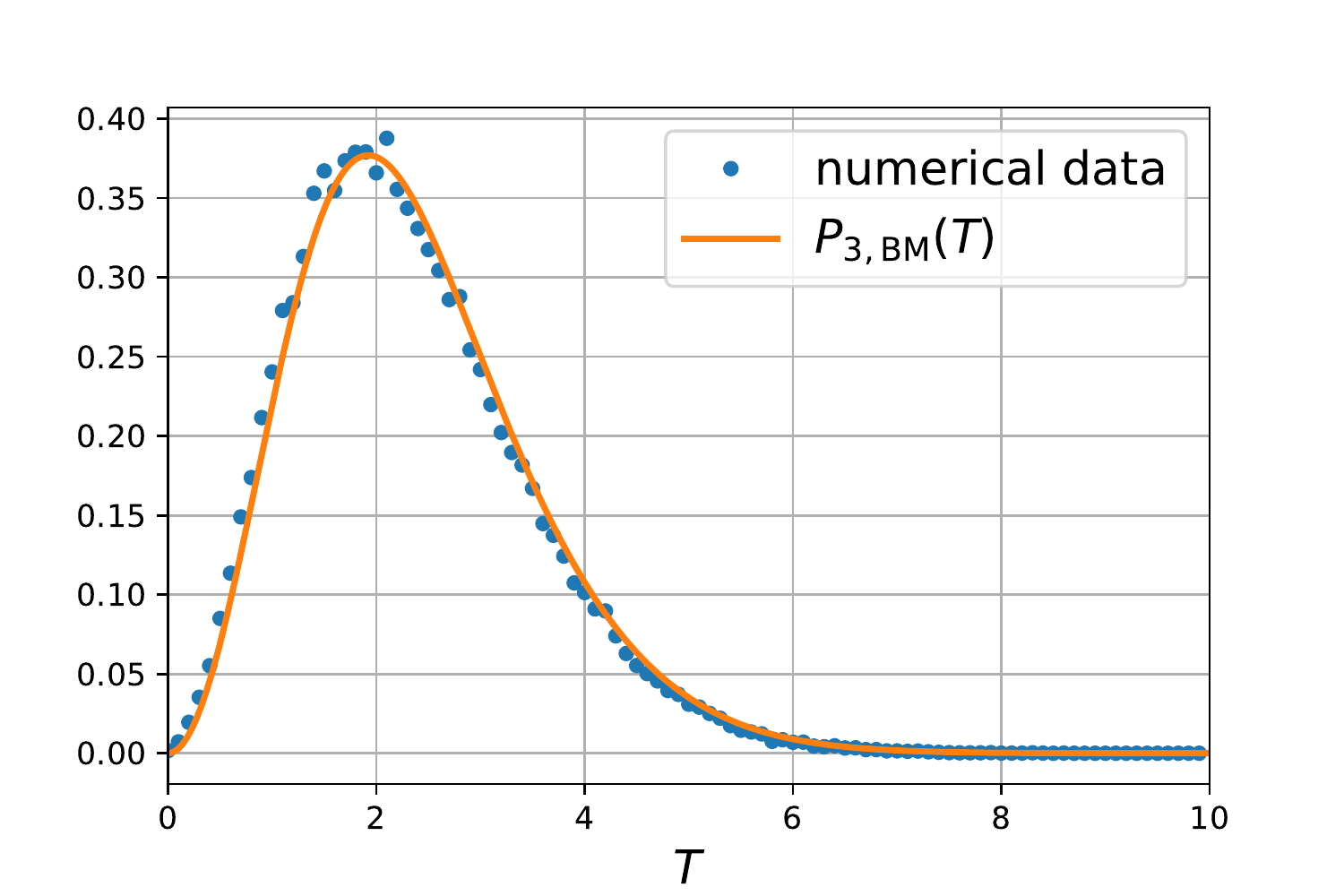}%
}
\subfloat{%
\includegraphics[width=0.49\textwidth]{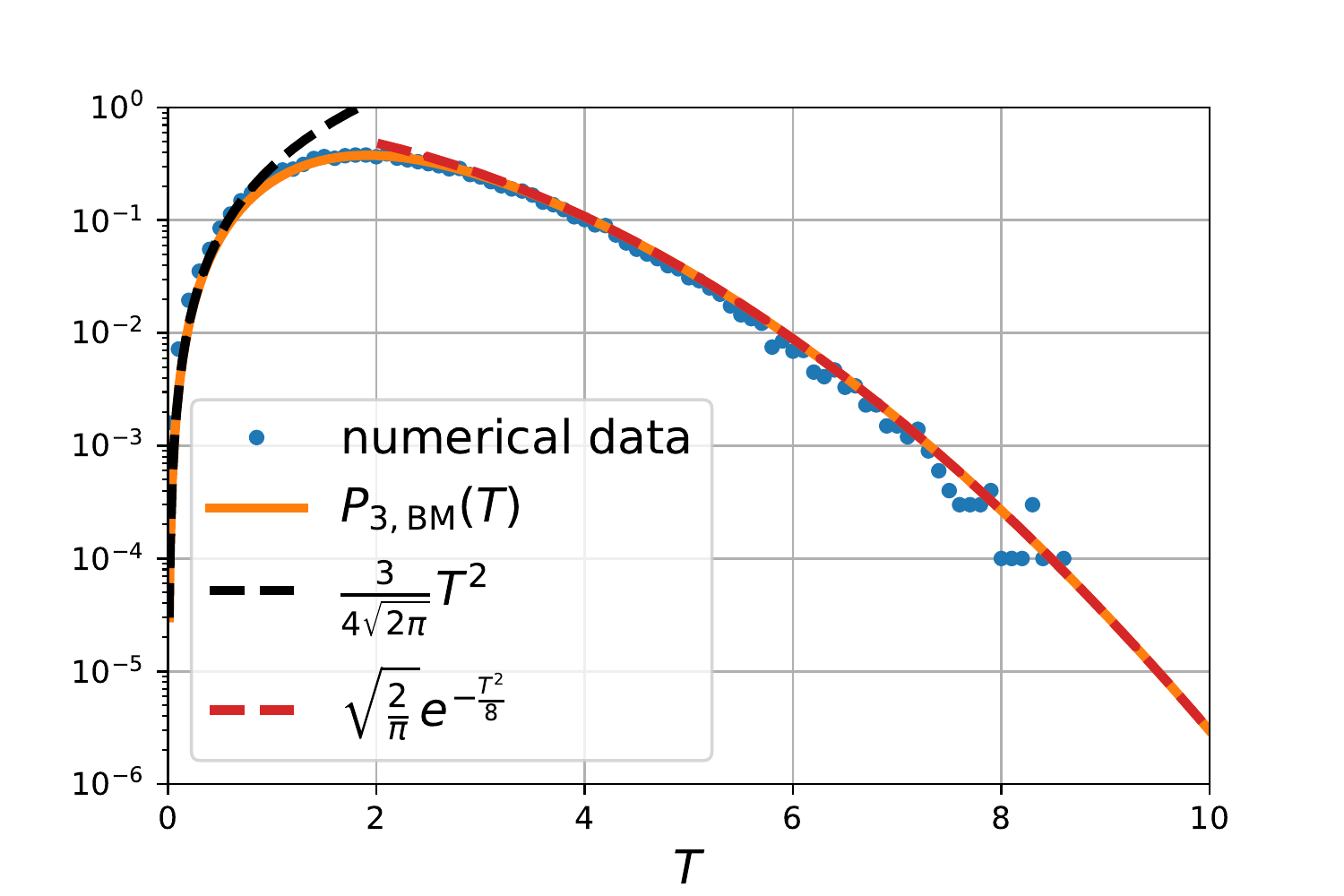}%
}
\caption{Comparison between the coincidence time obtained numerically for $N=3$ Brownian motions on the time interval $\tau \in[0,1]$ and diffusion coefficient $D=1$ and the analytical result in Eq. \eqref{P_3_BM_ana}. {\rm Left}: Linear scale, {\rm Right}: Logarithmic scale. }\label{P_3_BM}
\end{figure}

$\bm{N=3}$ The third moment gives the result for three walkers 
\be
\moy{e^{-c {\cal T}_3}}_{{\rm BM}}= \overline{  \mathcal Z_{\rm flat}(1)^3 } = 12 c \int_\mathbb{R} \frac{\rmd k}{2 \pi} \frac{k^2 e^{-2 k^2 }}{(k^2+c^2)(4 k^2 + c^2)}
\ee 
The inverse Laplace transform then yields
\begin{equation}
\begin{split}
P_{3,{\rm BM}}(T)=\mathcal{L}^{-1} \moy{e^{-c {\cal T}_3}}_{{\rm BM}} &=  \int_\mathbb{R} \frac{\rmd k}{2 \pi} 4 e^{-2 k^2 } (\cos(k T) - \cos(2 k T)) \\
&= \sqrt{\frac{2}{\pi}} e^{- \frac{T^2}{2}} (e^{\frac{3}{8} T^2} -1) \label{P_3_BM_ana}
\end{split}
\end{equation}
This result compares very well with the numerical simulations in Fig. \ref{P_3_BM}.

\subsubsection{Moment generating function for arbitrary $N$}
In the case of Brownian motions, the expression of the MGF of the coincidence time of $N$ walkers depends on the parity of $N$:\\

\textbf{For $N$ even} 
\bea\label{MGF_BM_even}
  \moy{e^{-c {\cal T}_N}}_{{\rm BM}}&& = \frac{(2 c)^{\frac{N}{2}}  N!}{(\frac{N}{2})!} \prod_{p=1}^{\frac{N}{2}} \int_\mathbb{R} \frac{\rmd k_p}{2 \pi}  \frac{e^{- 2 k_p^2  }}{4 k_p^2 + c^2}  \\
&& \times \prod_{1 \leqslant p<q \leqslant \frac{N}{2}} \frac{(k_p-k_q)^2}{(k_p-k_q)^2 + c^2} 
 \frac{(k_p+k_q)^2}{(k_p+k_q)^2 + c^2}  \nonumber 
\eea

\textbf{For $N$ odd} 
\bea\label{MGF_BM_odd}
  \moy{e^{-c {\cal T}_N}}_{{\rm BM}}&& =  \frac{(2 c)^{\frac{N-1}{2}}  N!}{(\frac{N-1}{2})!} \prod_{p=1}^{\frac{N-1}{2}} \int_\mathbb{R} \frac{\rmd k_p}{2 \pi} 
 \frac{k_p^2 e^{- 2 k_p^2 }}{(k_p^2+c^2)(4 k_p^2 + c^2)}  
\\
&& \times  
\prod_{1 \leqslant p<q \leqslant \frac{N-1}{2}} \frac{(k_p-k_q)^2}{(k_p-k_q)^2 + c^2} 
 \frac{(k_p+k_q)^2}{(k_p+k_q)^2 + c^2}  
 \nonumber
\eea 

These results are obtained from Eq. (108) in \cite{we-flatlong} by noting that for $c>0$ and
$N$ even only 
particle states with paired momenta $(k_1,-k_1,\dots,k_{N/2}, - k_{-N/2})$ contribute,
while for $N$ odd there are $N-1$ paired momenta and one zero momentum. 
\subsection{Probability Distribution Function of the coincidence time}
In this section, we obtain from Eqs.~\eqref{MGF_BM_even} and \eqref{MGF_BM_odd} an expression for the PDF of the coincidence time for $N$ Brownian motions.\\

\textbf{For $N$ even } For $p \in [1,N/2]$, define the conjugate variables
\begin{equation}
X_{2p}=c-2ik_p, \quad X_{2p-1}=c+2ik_p
\end{equation}
and observe that
\begin{equation}
\begin{split}
\frac{(k_p-k_q)^2}{(k_p-k_q)^2 + c^2} & \frac{(k_p+k_q)^2}{(k_p+k_q)^2 + c^2}  =\\
&\frac{(X_{2q}-X_{2p})(X_{2q-1}-X_{2p-1})(X_{2q-1}-X_{2p})(X_{2q}-X_{2p-1})}{(X_{2p}+X_{2q})(X_{2p-1}+X_{2q-1})(X_{2p-1}+X_{2q})(X_{2p}+X_{2q-1})}
\end{split}
\end{equation}
Then one rewrites Eq.~\eqref{MGF_BM_even} as
\be
  \moy{e^{-c {\cal T}_N}}_{{\rm BM}} = \frac{(2 c)^{\frac{N}{2}}  N!}{(\frac{N}{2})!} \prod_{p=1}^{\frac{N}{2}} \int_\mathbb{R} \frac{\rmd k_p}{2 \pi}  e^{- 2 k_p^2  }\prod_{p=1}^{\frac{N}{2}}\frac{1}{X_{2p}X_{2p-1}}\frac{X_{2p}+X_{2p-1}}{X_{2p}-X_{2p-1}}\prod_{1\leqslant p <q \leqslant N}\frac{X_q-X_p}{X_q+X_p}
\ee
Using the explicit values of $X_{2p}$ and $X_{2p-1}$ we simplify the expression as
\be
\label{eq:PDF_intermediate_BM_even}
  \moy{e^{-c {\cal T}_N}}_{{\rm BM}} = \frac{c^N  N!}{(\frac{N}{2})!} \prod_{p=1}^{\frac{N}{2}} \int_\mathbb{R} \frac{\rmd k_p}{2 \pi}  \frac{ie^{- 2 k_p^2  }}{k_p}\prod_{p=1}^{N}\frac{1}{X_{p}}\prod_{1\leqslant p <q \leqslant N}\frac{X_q-X_p}{X_q+X_p}
\ee
Using formulae (9.1) and (9.4) of Bruijn \cite{de1955some}, we write the last part of \eqref{eq:PDF_intermediate_BM_even} as
\begin{equation}
\prod_{p=1}^{N}\frac{1}{X_{p}}\prod_{1\leqslant p <q \leqslant N}\frac{X_q-X_p}{X_q+X_p}=\int_{\mathbb{R}_+^N}\rmd {\bf r}\; e^{-c\sum_\ell r_\ell} \prod_{1\leqslant k <\ell \leqslant N} \, {\rm sign}(r_k-r_\ell) \prod_{p=1}^{\frac{N}{2}} e^{-2ik_p(r_{2p-1}-r_{2p})  }
\end{equation}
From this identity, we obtain after anti-symmetrizing the last exponential w.r.t $r_{2p}$ and $r_{2p-1}$
\be
\begin{split}
&  \moy{e^{-c {\cal T}_N}}_{{\rm BM}} \\&= \frac{c^N  N!}{(\frac{N}{2})!} \int_{\mathbb{R}_+^N}\rmd {\bf r}\; e^{-c\sum_\ell r_\ell} \prod_{1\leqslant k <\ell \leqslant N} \, {\rm sign}(r_k-r_\ell) \prod_{p=1}^{\frac{N}{2}} \int_\mathbb{R} \frac{\rmd k_p}{2 \pi}  e^{- 2 k_p^2}\frac{\sin(2k_p(r_{2p-1}-r_{2p}))}{k_p}\\
&= \frac{c^N  N!}{2^{\frac{N}{2}}(\frac{N}{2})!} \int_{\mathbb{R}_+^N}\rmd {\bf r}\; e^{-c\sum_\ell r_\ell} \prod_{1\leqslant k <\ell \leqslant N} \, {\rm sign}(r_k-r_\ell) \prod_{p=1}^{\frac{N}{2}}\, \erf\left(\frac{r_{2p-1}-r_{2p}}{\sqrt{2}}\right)
\end{split}
\ee
where we computed the integral w.r.t $\lbrace k_p \rbrace$'s. The final trick consists in \textit{(i)} changing the labels in the $r$ variables in the error function from $r_{2p}$ to $r_{\sigma(2p)}$ where $\sigma$ belongs to the symmetric group $S_N$, \textit{(ii)} using the fact that there are $N!/(2^{N/2}(N/2)!)$ ways of pairing $N$ objects ($N$ is even) and \textit{(iii)} using the definition of the Pfaffian of an anti-symmetric matrix ${\rm Pf}(A)=\sum_{\sigma\in S_N, \sigma(2p-1)<\sigma(2p)}{\rm sign}(\sigma)\prod_{p=1}^{N/2}A_{\sigma(2p-1),\sigma(2p)}$ to finally obtain
\be
  \moy{e^{-c {\cal T}_N}}_{{\rm BM}} =c^N \int_{\mathbb{R}_+^N}\rmd {\bf r}\; e^{-c\sum_\ell r_\ell} \prod_{1\leqslant k <\ell \leqslant N} \, {\rm sign}(r_k-r_\ell)\;  \underset{1\leqslant k,\ell\leqslant N}{\rm Pf}\left[\, \erf\left(\frac{r_{k}-r_{\ell}}{\sqrt{2}}\right)\right]
\ee

Inverting the Laplace transform of this expression, we obtain the PDF $P_{N,{\rm BM}}(T)$ of the rescaled coincidence time ${\cal T}_N= {\cal T}_N(t=1)$ as
\begin{align}
&P_{N,{\rm BM}}(T)=\partial_T^N \int_{\mathbb{R}_+^{N}} \rmd {\bf r}\; \delta\left(T-\sum_{\ell=1}^N  r_\ell\right) \prod_{1\leqslant k <\ell \leqslant N} \, {\rm sign}(r_k-r_\ell)\;  \underset{1\leqslant k,\ell\leqslant N}{\rm Pf}\left[\, \erf\left(\frac{r_{k}-r_{\ell}}{\sqrt{2}}\right)\right]\nn\\
&=\partial_T^{N+1} \int_{\mathbb{R}_+^{N}} \rmd {\bf r}\; \Theta\left(T-\sum_{\ell=1}^N r_\ell\right)\prod_{1\leqslant k <\ell \leqslant N} \, {\rm sign}(r_k-r_\ell)\;  \underset{1\leqslant k,\ell\leqslant N}{\rm Pf}\left[\, \erf\left(\frac{r_{k}-r_{\ell}}{\sqrt{2}}\right)\right]\label{PDF_P_BM}
\end{align}
where $\Theta(x)$ is the Heaviside step function. Even though the product of sign functions can itself be written as a Pfaffian, we have not tried to simplify further Eq.~\eqref{PDF_P_BM}.\\

\textbf{For $N$ odd}  
For $p \in [1,(N-1)/2]$, define the conjugate variables
\begin{equation}
X_{2p}=c-2ik_p, \quad X_{2p-1}=c+2ik_p
\end{equation}
and define $X_N=c$. Then, in a similar fashion as the even $N$ case, one rewrites Eq.~\eqref{MGF_BM_odd} as

\be
  \moy{e^{-c {\cal T}_N}}_{{\rm BM}} =  \frac{c^N   N!}{(\frac{N-1}{2})!} \prod_{p=1}^{\frac{N-1}{2}} \int_\mathbb{R} \frac{\rmd k_p}{2 \pi} 
\frac{ie^{- 2 k_p^2 }}{k_p}\prod_{p=1}^N \frac{1}{X_p}\prod_{1\leqslant p <q \leqslant N}\frac{X_q-X_p}{X_q+X_p} 
\ee
By the same argument as in the even $N$  case, using Bruijn's results  \cite{de1955some}, the moment generating function reads
\be
  \moy{e^{-c {\cal T}_N}}_{{\rm BM}} =  \frac{c^N   N!}{2^{\frac{N-1}{2}}(\frac{N-1}{2})!}\int_{\mathbb{R}_+^N}\rmd {\bf r}\; e^{-c\sum_\ell r_\ell} \prod_{1\leqslant k <\ell \leqslant N} \, {\rm sign}(r_k-r_\ell) \prod_{p=1}^{\frac{N-1}{2}} \, \erf\left(\frac{r_{2p-1}-r_{2p}}{\sqrt{2}}\right)
\ee
We have to be careful in the odd case as there are $N$ variables $r$ but only $N-1$ of them are involved in the error functions whereas all $N$ variables are involved in the product of sign functions. We now wish to employ the same Pfaffian trick as the even $N$ case and hence have to consider all $(N-1)!/(2^{\frac{N-1}{2}}(\frac{N-1}{2})!)$ ways of pairings $N-1$ objects. Hence, we obtain our final expression for the moment generating function in the odd $N$ case
\be
  \moy{e^{-c {\cal T}_N}}_{{\rm BM}} =N c^N \int_{\mathbb{R}_+^N}\rmd {\bf r}\; e^{-c\sum_\ell r_\ell} \prod_{1\leqslant k <\ell \leqslant N} \, {\rm sign}(r_k-r_\ell)\;  \underset{1\leqslant k,\ell\leqslant N-1}{\rm Pf}\left[\, \erf\left(\frac{r_{k}-r_{\ell}}{\sqrt{2}}\right)\right]
\ee
Inverting the Laplace transform of this expression, we obtain the PDF $P_{N,{\rm BM}}(T)$ of the rescaled coincidence time ${\cal T}_N= {\cal T}_N(t=1)$ as
\begin{align}
&P_{N,{\rm BM}}(T)\\
&=N\partial_T^N \int_{\mathbb{R}_+^{N}} \rmd {\bf r}\; \delta\left(T-\sum_{\ell=1}^N r_\ell\right) \prod_{1\leqslant k <\ell \leqslant N} \, {\rm sign}(r_k-r_\ell)\;  \underset{1\leqslant k,\ell\leqslant N-1}{\rm Pf}\left[\, \erf\left(\frac{r_{k}-r_{\ell}}{\sqrt{2}}\right)\right]\nn\\
&=N\partial_T^{N+1} \int_{\mathbb{R}_+^{N}} \rmd {\bf r}\; \Theta\left(T-\sum_{\ell=1}^N r_\ell\right)\prod_{1\leqslant k <\ell \leqslant N} \, {\rm sign}(r_k-r_\ell)\;  \underset{1\leqslant k,\ell\leqslant N-1}{\rm Pf}\left[\, \erf\left(\frac{r_{k}-r_{\ell}}{\sqrt{2}}\right)\right]\label{PDF_P_BM_odd}
\end{align}
where $\Theta(x)$ is the Heaviside step function. We have verified that for $N=2,3$, these formulae give back the results for the PDF and the MGF for the coincidence time of Brownian motions.
For $N=2,3$ we only have to use a $2 \times 2$ Pfaffian, $\underset{1\leqslant i,j\leqslant 2}{\rm Pf} A_{ij}=A_{12}$,
hence the calculation is elementary.

\subsection{Asymptotic behaviour of $P_{N,{\rm BM}}(T)$ for arbitrary $N$}\label{BM_asy}
\subsubsection{Small $T$ limit of the Probability Distribution Function}
We now discuss the small $T$ behavior of the PDF $P_{N,{\rm BM}}(T)$ of ${\cal T}_N={\cal T}_N(t)/\sqrt{t}$. 
It can be extracted from  the $c\to +\infty$ limit of $\moy{e^{-c {\cal T}_N}}_{{\rm BM}}$.  When $c$ is increased, the interaction in the Lieb-Liniger model
becomes more repulsive and the corresponding Brownian trajectories 
with small coincidence time are those where
none of the Brownian are bounded together.
Taking the large $c\to +\infty$ limit of Eqs. \eqref{MGF_BM_even} and \eqref{MGF_BM_odd} 
we see that the leading term is in all cases $c^{-N(N-1)/2}$.   Inverting the Laplace transform, we obtain the small $T$ behavior of $P_{N,{\rm BM}}(T) $ as
\be \label{smallTflat1}
P_{N,{\rm BM}}(T) \underset{T\to 0}{=} I_N  T^{\frac{N(N-1)}{2}-1}+\mathcal{O}(T^{\frac{N(N-1)}{2}+1}),
\ee
where $I_N$ depends on the parity of $N$.\\

\textbf{For $N$ even} 

\begin{align} 
I_{N}&=\frac{2^{\frac{N}{2}}N!}{(\frac{N}{2})!\Gamma\left(\frac{N(N-1)}{2}\right)} \prod_{p=1}^{\frac{N}{2}} \int_\mathbb{R} \frac{\rmd k_p}{2 \pi}e^{- 2 k_p^2  } 
\prod_{1 \leqslant p<q \leqslant \frac{N}{2}} (k_p^2-k_q^2)^2\nn\\
&=\frac{\Gamma(N+1)}{2^{\frac{N(N-2)}{2}}(2\pi)^{\frac{N}{4}}\Gamma\left(\frac{N(N-1)}{2}\right)}\prod_{k=1}^{\frac{N}{2}-1}\Gamma(2k+1)\label{I_even}
\end{align}

\textbf{For $N$ odd} 

\begin{align}
I_{N}&=\frac{2^{\frac{N-1}{2}}  N!}{(\frac{N-1}{2})!\Gamma\left(\frac{N(N-1)}{2}\right)}\prod_{p=1}^{\frac{N-1}{2}} \int_\mathbb{R} \frac{\rmd k_p}{2 \pi}k_p^2 e^{- 2 k_p^2 } \prod_{1 \leqslant p<q \leqslant \frac{N-1}{2}}(k_p^2-k_q^2)^2\nn\\
&=\frac{\Gamma(\frac{N}{2}+1)}{2^{\frac{N(N-3)}{2}}(2\pi)^{\frac{N+1}{4}}\Gamma\left(\frac{N(N-1)}{2}\right)}\prod_{k=1}^{\frac{N-1}{2}}\Gamma(2k+1)\;.\label{I_odd}
\end{align}

The $k$ integrals are typical examples of Selberg integrals, 
and further details can be found in Ref.~\cite{FoWa08} in Section 1.4 and after changing variables from $\lbrace x_i\rbrace $'s to $\lbrace k_i=x_i/\sqrt{2} \rbrace$. One can check that these results agree with the small $T$ expansion of the 
formula for the PDF in the cases $N=2,3$ given above, i.e. $I_2=\sqrt{\frac{2}{\pi}}$ and 
$I_3= \frac{3}{4 \sqrt{2 \pi}}$.

\subsubsection{Large $T$ limit of the Probability Distribution Function}

Similarly to Section~\ref{subsec:largeT_BB}, we determine the large $T$ tail of the PDF $P_{N,{\rm BB}}(T)$ of ${\cal T}_N={\cal T}_N(t)/\sqrt{t}$ by investigating the $c\to -\infty$ limit of the Lieb-Liniger model. It again
determined by the same ground state as in Section~\ref{subsec:largeT_BB}. From Eqs.~(65-66) in  
Ref.~\cite{KPZLateTime} it follows that the MGF in the $c \to - \infty$ limit reads
\begin{equation}
\moy{e^{-c {\cal T}_N}}_{{\rm BM}}=2^{N-1} e^{{\bar c}^2 \frac{N(N^2-1)}{12}}[1+\mathcal{O}(e^{-\frac{1}{4}N(N-1)\bar{c}^2})]
\end{equation}
As in Section~\ref{subsec:largeT_BB} we find the $T\to \infty$ asymptotics
\begin{equation}
P_{N,\rm{BM}}(T)=2^{N-1}\sqrt{\frac{3}{\pi N(N^2-1)}}  e^{-\alpha_N T^2}+\mathcal{O}(e^{-\beta_N T^2})
\end{equation}
with the exponential factors again given by eqs.~\eqref{eq:largeT_expFactor}, \textit{i.e.} $\alpha_N=\frac{3}{N(N^2-1)}$ and $\beta_N=\frac{3}{N^3-3N^2+2N}$.
For completeness, we compute this asymptotics explicitly for $N=2,3$.
\begin{itemize}
\item $N=2$
\begin{equation}
P_{2,\rm{BM}}(T)= \sqrt{\frac{2}{\pi }} e^{-\frac{T^2}{2}}
\end{equation}
This matches eq.~\eqref{P_2_BM_ana}.
\item $N=3$
\begin{equation}
P_{3,\rm{BM}}(T)=\sqrt{\frac{2}{\pi }} e^{-\frac{T^2}{8}}+\mathcal{O}(e^{-\frac{T^2}{2}})
\end{equation}
This matches eq.~\eqref{P_3_BM_ana}.
\end{itemize}

\subsection{Mean and variance of the coincidence time for $N$ Brownian motions}
~~~
\textbf{Mean value of ${\cal T}_{N,{\rm BM}}$} To compute the first moment, we may use the propagator for a single Brownian motion with diffusion coefficient $D$
\be
P(x,\tau|x_0,0)=\frac{e^{-\frac{(x-x_0)^2}{4D\tau}}}{\sqrt{4\pi D\tau}}\;,
\ee
together with the fact that we consider i.i.d. variables. 
The mean value of the coincidence time of the Brownian motion can be obtained (setting $D=1$) as
\begin{align}
\moy{{\cal T}_N}_{\rm BM}&=\int_0^{1}\rmd \tau \sum_{i\neq j} \moy{\delta(x_i(\tau)-x_j(\tau))}=\int_0
^{1}\rmd \tau \sum_{i\neq j} \int_{-\infty}^{\infty} P(x_i,\tau|x_0,0)^2 \rmd x_i\nn\\
&=\frac{N(N-1)}{2}\int_0^{1} \frac{\rmd \tau}{\sqrt{\tau}} \int_{-\infty}^{\infty}\frac{\rmd x}{4\pi}e^{-\frac{x^2}{2}}=\frac{N(N-1)}{2}\sqrt{\frac{2}{\pi}}\;.\label{mean_BM}
\end{align}
As expected, we have $\moy{{\cal T}_N}_{\rm BM}<\moy{{\cal T}_N}_{\rm BB}$.\\

\textbf{Variance of ${\cal T}_{N,{\rm BM}}$} We may still use the free propagator to compute the second moment of the distribution. Using again the Brownian propagator, we obtain
\begin{align}
&\moy{{\cal T}_N^2}=\int_0^{1}\rmd \tau_1 \int_0^{1}\rmd \tau_2 \sum_{i\neq j}\sum_{\ell\neq m} \moy{\delta(x_i(\tau_1)-x_j(\tau_1))\delta(x_m(\tau_2)-x_\ell(\tau_2))}\\
&=\frac{N!}{(N-4)!}\left[\int_0^{1}\rmd \tau \int_{-\infty}^{\infty} P(x,\tau|0,0)^2 \rmd x\right]^2 \nn\\
&+8\frac{ N!}{(N-3)!}\int_0^{1}\rmd \tau_1 \int_{\tau_1}^{1}\rmd \tau_2  \int_{-\infty}^{\infty}\int_{-\infty}^{\infty} P(x,\tau_1|0,0)^2 P(y-x,\tau_2-\tau_1|0,0) \nn\\
& \hspace*{7cm} \times  P(y,\tau_2|0,0) \rmd x\rmd y \nn\\
&+4x\frac{ N!}{(N-2)!}\int_0^{1}\rmd \tau_1 \int_{\tau_1}^{1}\rmd \tau_2  \int_{-\infty}^{\infty}\int_{-\infty}^{\infty} P(x,\tau_1|0,0)^2 P(y-x,\tau_2-\tau_1|0,0)^2 \rmd x\rmd y\;. \nn
\end{align}
After a careful computation, we finally obtain 
\be
\var{{\cal T}_N}_{\rm BM}=\moy{\left[{\cal T}_N-\moy{{\cal T}_N}_{\rm BM}\right]^2}_{\rm BM}=\frac{N!}{(N-3)!}\left[\frac{2}{3}-\frac{2}{\pi}\right]+\frac{N!}{(N-2)!}\left[\frac{1}{2}-\frac{1}{\pi}\right]\;.\label{var_BM}
\ee
We note that the variance ratio $\var{{\cal T}_N}_{\rm BM}/\var{{\cal T}_N}_{\rm BB}$
decreases monotonically from $0.846638\dots$ for $N=2$ to
$0.724549\dots$ for infinite $N$. The ratio of relative variances
$\moy{{\cal T}_N}^2_{\rm BB} \var{{\cal T}_N}_{\rm BM}/(\var{{\cal T}_N}_{\rm BB} \moy{{\cal T}_N}^2_{\rm BM})$ 
decreases from $2.088996\dots$ for $N=2$ to $1.7877536\dots$ for infinite $N$.

\section{Coincidence time for arbitrary fixed final points}\label{sec:coincidence}

As mentionned in Section \ref{subsec:Bethe} there is an alternative formula 
\eqref{C_MGF} to the one from the Bethe ansatz, for the moments of the directed polymer problem, obtained
from the study of Macdonald processes. It turns
out that this formula can be extended to arbitrary final points ${\bf y}$ \cite{BorodinMacdo,G18}
(while the starting points are still all at zero, i.e. ${\bf x}_0=0$) and for arbitrary values of $c$.
From this formula, specialized to the case $c>0$, we immediately 
obtain that for $y_1 \leqslant y_2 \leqslant \dots \leqslant y_N$, the 
Laplace transform of the PDF of the coincidence time for $N$ such Brownian walkers is 
\be\label{C_MGF_ap}
\moy{e^{-c {\cal T}_N(t)}}_{{\bf 0},{\bf y}}=\left(4 \pi t \right)^{\frac{N}{2}} \int_{i\mathbb{R}^N}\frac{\rmd {\bf z}}{(2 i \pi)^N} e^{t ({\bf z} + \frac{{\bf y}}{2 t})^2} \prod_{i<j}\frac{z_i-z_j}{z_i-z_j+c}\;.
\ee

We note that the r.h.s. of 
\eqref{C_MGF_ap} is invariant by a global shift of the endpoints $y_i \to y_i + w$. 
This global shift corresponds to adding a linear drift to each of the Brownians
which has no effect on the coincidence time.

\subsection{Exact distribution for $N=2,3$}
~
\textbf{Distribution for $N=2$.} We start from Eq. \eqref{C_MGF_ap} where we have set $N=2$ and $t=1$. This yields
\be
\moy{e^{-c {\cal T}_2}}_{{\bf 0},{\bf y}}= 4 \pi \int_{i\mathbb{R}^2}\frac{\rmd z_1\rmd z_2}{(2 i \pi)^2} e^{ (z_1 + \frac{y_1}{2})^2+(z_2 + \frac{y_2}{2})^2}\frac{z_1-z_2}{z_1-z_2+c}\;,\;\;y_1\leqslant y_2\;.
\ee
Taking the inverse Laplace transform and introducing the change of variables $z_j\to i k_j$
one obtains

\begin{align}
P_2(T,y_1,y_2)&= 4 \pi \int_{\mathbb{R}^2}\frac{\rmd k_1\rmd k_2}{(2\pi)^2}(i k_1-i k_2) e^{ -(k_1 - \frac{i y_1}{2})^2-(k_2 - \frac{i y_2}{2})^2-T(i k_1- i k_2)}\nn\\
&=-\partial_T\int_{\mathbb{R}^2}\frac{\rmd k_1\rmd k_2}{\pi}e^{ -(k_1 - \frac{i y_1}{2})^2-(k_2 - \frac{i y_2}{2})^2-iT(k_1-k_2)}\;.
\end{align}
The two Gaussian integrals can now be computed 
and one obtains
\be
P_2(T,y_1,y_2)=\frac{1}{2}(2T+y_2-y_1)e^{-\frac{T}{2}(T+y_2-y_1)}\;,\;\;y_1\leqslant y_2\;.
\ee
Note that setting $y_2=y_1$, we recover the result for the Brownian bridge \eqref{P_2_BB_ana}.
Alternatively, the result for the Brownian motion \eqref{P_2_BM_ana} is obtained by computing
\be
P_{2,{\rm BM}}(T)=2 \int_{-\infty}^{\infty}\rmd y_1\int_{y_1}^{\infty}\rmd y_2\frac{e^{-\frac{y_1^2+y_2^2}{4}}}{4\pi}P_2(T,y_1,y_2)\;.
\ee
Finally, symmetrising the problem by releasing the constraint $y_1\leqslant y_2$, by defining 
\be
P_{2,{\rm sym}}(T,y_1,y_2)=\frac{1}{2}(2T+|y_2-y_1|)e^{-\frac{T}{2}(T+|y_2-y_1|)}\;,
\ee
we may obtain the joint PDF of the coincidence time, the algebraic distance $d$ between the Brownian defined as $d=y_1-y_2$ (with $d\in\mathbb{R}$)  and their center of mass position $u=(y_1+y_2)/2$. It reads 
\be
\begin{split}
\tilde {P_2}(T, d,u)&=\frac{e^{-\frac{\left(\frac{d}{2}+u\right)^2+\left(-\frac{d}{2}+u\right)^2}{4}}}{4\pi} P_{2,{\rm sym}}\left(T,-\frac{d}{2}+u,\frac{d}{2}+u\right)\\
&=\frac{e^{-\frac{u^2}{2}}}{8\pi}(2T+|d|)e^{-\frac{1}{8}(2T+|d|)^2}\;.
\end{split}
\ee
We may then recover the joint PDF of $d \in \mathbb{R}$ and $T\in \mathbb{R}_+$ in Eq. \eqref{P_joint_T_d} by integrating over the center of mass position 
$u$,
\be
P_{\rm joint}(T,d)=\int_{-\infty}^{\infty}\tilde {P_2}(T, d,u)\rmd u=\frac{(2T+|d|)}{2\sqrt{8\pi}}e^{-\frac{1}{8}(2T+|d|)^2}
\ee

\textbf{Distribution for $N=3$.}  The distribution for $N=3$ can be obtained from Eq. \eqref{C_MGF_ap} but its expression is quite cumbersome and not very enlightening. Therefore, we only reproduce here its asymptotic behaviours for $y_1<y_2<y_3$
\be
P_{3}(T,y_1,y_2,y_3)=\begin{cases}
\displaystyle  \frac{(y_2-y_1)(y_3-y_2)(y_3-y_1)}{16}T^2+\mathcal{O}(T^3) &\;,\;\;T\to 0\\
&\\
\displaystyle \frac{1}{4}\sqrt{\frac{\pi}{6}} e^{\frac{(y_1+y_3-2y_2)^2}{24}-\frac{T}{4}(y_3-y_1)-\frac{T^2}{8}}\left[\left(T+y_3-y_1\right)^2-4\right]&\;,\;\;T\to \infty\;.
\end{cases}
\ee
Note that taking two equal points $y_2=y_3$, the PDF behaves for small $T$ as $P_{3}(T,y_1,y_2,y_2)=(y_1-y_2)^2 T^3/24+\mathcal{O}(T^4)$, while for all equal points $y_1=y_2=y_3$, we recover the expression in Eq. \eqref{eq:P3BB_asympt}.

\subsection{Small $T$ asymptotics of the PDF for any $N$ and for any fixed final points}

From \eqref{C_MGF_ap} one can extract the small $T$ asymptotics for any $N$ and any fixed final points
${\bf y}$. In the large $c$ limit it becomes
\bea \label{eq:large_C_anyEnd}
 \moy{e^{-c {\cal T}_N}}_{{\bf 0},{\bf y}} &\simeq& \left(4 \pi \right)^{\frac{N}{2}} 
c^{- \frac{N(N-1)}{2}} 
\int_{i\mathbb{R}^N}\frac{\rmd {\bf z}}{(2 i \pi)^N} e^{({\bf z} + \frac{{\bf y}}{2})^2} \prod_{i<j} (z_i-z_j)\;. \\
& = &\left(4 \pi \right)^{\frac{N}{2}} 
c^{- \frac{N(N-1)}{2}} 
\int_{i\mathbb{R}^N}\frac{\rmd {\bf z}}{(2 i \pi)^N} e^{{\bf z}^2} \prod_{i<j} (-\frac{y_i}{2} +\frac{y_j}{2}  )\;. \nonumber 
\\
& = &  (2 c)^{- \frac{N(N-1)}{2}} \prod_{i<j} (y_j - y_i)
\eea
The second line follows from the fact that the first line is antisymmetric in the $y_i$ and upon
the shift $z_i \to z_i - y_i/2$, that is is a polynomial of degree $N(N-1)/2$ in the $y_i$. Taken together these properties
imply that it is proportional to the Vandermonde product $\prod_{i<j} (y_j - y_i)$. Upon inverse Laplace
inversion we obtain
\be \label{Py} 
P_N(T,{\bf y}) \simeq \frac{1}{2^{\frac{N(N-1)}{2}} \Gamma(\frac{N(N-1)}{2})} \prod_{i<j} |y_j - y_i| ~ T^{\frac{N(N-1)}{2}-1}  
\ee
where we use that it must be a symmetric function of the endpoints. It agrees with the above results for
$N=2,3$. The 
results \eqref{smallTflat1}, \eqref{I_even}, \eqref{I_odd}
for the BM is recovered from the identity 
\be
P_{N,{\rm BM}}(T) = \int_{\mathbb{R}^N} \frac{\rmd {\bf y}}{(4 \pi)^{N/2}} e^{- \sum_i \frac{y_i^2}{4} } P_N(T,{\bf y})
\ee
Using again the Mehta integral formula e.g. (1.5)-(1.6) in Ref.~\cite{FoWa08} we
recover \eqref{smallTflat1} with an amplitude
\be \label{INbis} 
I_N =  \frac{1}{2^{\frac{N(N-1)}{4}} \Gamma(\frac{N(N-1)}{2})} (\frac{2}{\sqrt{\pi}})^N
\prod_{j=1}^N \Gamma(1+ \frac{j}{2}) 
\ee
One checks that this (simpler) expression for $I_N$ is equivalent to
those in \eqref{I_even}, \eqref{I_odd} obtained by a quite different method. To prove the equivalence, one takes Eqs.~\eqref{I_even} and \eqref{I_odd} and insert the duplication formula $\Gamma(2k+1)=\Gamma(k+\frac{1}{2})\Gamma(k+1)2^{2k}/\sqrt{\pi}$ and the identity \eqref{INbis} follows.

\subsection{Large $T$ asymptotics of the PDF for any $N$ and arbitrary final {\it and} initial points}
A formula for $P_N(T)$ for any $N$ and arbitrary initial and final points seems out of reach at present. Indeed it would require to handle all eigenstates of the Lieb-Liniger Hamiltonian 
\eqref{H_LL}, i.e. with arbitrary symmetry, and not just the fully symmetric (\textit{i.e.} bosonic) ones.
For attempts at solving this problem in the KPZ/directed polymer context
using the so-called generalized Bethe ansatz see Refs.~\cite{CNYang_bethe,BSutherland_bethe,EmigKardar,DeLucaPLD1,DeLucaPLD2,DeLucaPLD3}.\\

It is however possible to extract the leading asymptotics at large $T$ for arbitrary 
fixed final and initial points. Indeed, as discussed above, this asymptotics 
is controled in the MGF by large negative values $c=-\bar c<0$. In that limit,
all moments are dominated by the ground state of the Lieb-Liniger Hamiltonian \eqref{H_LL}.
The key fact is that this ground state is the bosonic one which we used before, where
all particles are bound in a single string. Eigenstates with other symmetries have a higher energy (and at zero total momentum, differ by a finite gap from the ground state). To treat the case of arbitrary endpoints we now use the known form of the ground state eigenfunction 
\be
\Psi_{0,k}(x) = N! e^{- \frac{\bar c}{2} \sum_{1 \leqslant i<j \leqslant N} |x_i-x_j| + i k \sum_{j=1}^N x_j}
\ee
where $k$ is the momentum of the center of mass of the string. Its
energies are $E_{0,k}(N)=- c^2 \frac{N(N^2-1)}{12} + N k^2$. 
Note that we must sum over all values of $k$, i.e. it is a 
ground state manifold, but this is already what we did above to obtain 
the large $T$ asymptotics for the BB and the BM. 
We can now write, in the large $\bar c$ limit
\be
Z_N({\bf x},t|{\bf y};c)=\langle {\bf x}|e^{-\hat{\cal H}_N(c) t}|{\bf y} \rangle
\simeq  e^{\frac{c^2}{12} N(N^2-1) t} \int_\mathbb{R} \frac{\rmd k}{2 \pi} \frac{N L e^{-N k^2 t}}{||\Psi_{0,k}||^2} \Psi_{0,k}({\bf x}) \Psi^*_{0,k}({\bf y}) 
\ee
where the norm is 
$||\Psi_{0,k}||^2=  N! N^2 \bar c^{1-N} L$. As in Ref.~\cite{CDR10},
we keep the leading order in large $L$, but the factors of $L$
cancel in the final result. This leads to the moment generating function

\bea
\fl && \moy{e^{\bar c {\cal T}_N}}_{{\bf 0},{\bf y}} 
= \frac{Z_N({\bf x},1|{\bf y};c)}{Z_N({\bf x},1|{\bf y};c=0)}
= (4 \pi)^{\frac{N}{2}} e^{ \frac{1}{4} \sum_i (x_i-y_i)^2 } Z_N({\bf x},1|{\bf y};c) \\
\fl && 
\simeq (4 \pi)^{\frac{N}{2}} \bar c^{N-1} N!  e^{\frac{N^3-N}{12} \bar c^2} 
e^{- \frac{\bar c}{2} \sum_{i<j} ( |y_i-y_j| + |x_i-x_j| )}
\int_\mathbb{R} \frac{\rmd k e^{- N k^2} }{2 \pi N}  
 e^{\sum_j( i k  (x_j-y_j)
 + \frac{(x_j-y_j)^2}{4} ) } \nonumber \\
\fl && = (4 \pi)^{\frac{N-1}{2}} \bar c^{N-1} N!  N^{-\frac{3}{2}} e^{\frac{N^3-N}{12} \bar c^2} 
e^{- \frac{\bar c}{2} \sum_{i<j} ( |y_i-y_j| + |x_i-x_j| )}
e^{- \frac{1}{4 N} [\sum_j (x_j-y_j)]^2 
 +\sum_j  \frac{(x_j-y_j)^2}{4} } \nonumber
\eea

From this we obtain the following leading large $T$ asymptotics for the PDF of the
coincidence time with fixed initial and final points
\bea \label{anyXY} 
\fl && P_{N}(T,{\bf x},{\bf y}) \simeq \frac{N!\,2^{N-1}\pi^{\frac{N}{2}-1} \sqrt{\alpha_N} }{N^{3/2}} 
\,e^{ \sum_{i=1}^N \frac{(y_i - \bar y - x_i + \bar x)^2}{4} } \, e^{- \frac{\alpha_N}{4} ({\sf d}[{\bf y}] + {\sf d}[{\bf x}])^2} \\
\fl && ~~~~~~~~~~~~~~~~~~~~~~~~~~~~~~~~~~~~~~~~~~~~~~~~~~~~~~
  \times (-\partial_T)^{N-1}  e^{- \alpha_N  ({\sf d}[{\bf y}] + {\sf d}[{\bf x}]) T -\alpha_N T^2} \nonumber \\ \hspace*{1em}
\fl && {\sf d}[{\bf y}] := \sum_{1 \leqslant i<j \leqslant N} |y_i-y_j| \quad , \quad 
\bar y := \frac{1}{N} \sum_{i=1}^N y_i \quad , \quad \alpha_N=\frac{3}{N(N^2-1)}
\eea
Hence to this order in the large $T$ expansion, the PDF depends only on \textit{(i)} the sum of the \textit{total distance} of the final points, ${\sf d}[{\bf y}]$,
and of the initial points, ${\sf d}[{\bf x}]$ and 
\textit{(ii)} the sum of the squares of the centered variables $y_i-x_i - (\bar y - \bar x)$.\\

We thus note that the PDF is invariant by the global shift $y_i-x_i \to y_i-x_i + w$, 
an exact property, not restricted to the tail, as previously discussed. We also note that
the PDF is symmetric in the simultaneous exchange of all initial and final points ${\bf y} \leftrightarrow {\bf x}$,
which again should be an exact property from the time reversibility of Brownian motion.

\subsection{Mean coincidence time for arbitrary final points}

We can extend the calculation of Section \ref{BB_mean_var} using formula \eqref{C_MGF_ap}
for arbitrary final points ${\bf y}$. The generalization being straightforward we give only a few steps.
One finds for $y_1 \leqslant y_2 \leqslant \dots \leqslant y_N$.
\bea
\fl  \moy{{\cal T}_N}_{{\bf 0},{\bf y}} &=&  4\pi 
\sum_{i<j} \int_0^{\infty}\rmd r\int_{\mathbb{R}^2}
\frac{\rmd k_i \rmd k_j }{(2  \pi)^2} e^{-(k_i- i \frac{y_i}{2})^2-(k_j- i \frac{y_j}{2})^2}e^{-i(k_i-k_j)r-0^{+}r} \\
\fl & =& \sum_{i<j} \int_0^{\infty}\rmd r e^{-\frac{1}{2} r (r+y_j-y_i)} 
= \sqrt{\frac{\pi }{2}} \sum_{i<j}  e^{\frac{1}{8} (y_i-y_j)^2}
   \left(1-\text{erf}\left(\frac{y_j-y_i}{2 \sqrt{2}}\right) \right)
\eea
Note that it decays as $\moy{{\cal T}_N}_{{\bf 0},{\bf y}}  \sim  \sum_{i<j} \frac2{y_j-y_i}$
when the final points are all taken very far apart. 

\section{Conclusion}

We have investigated the probability distribution function $P_N(T)$ of the total pairwise coincidence time ${\cal T}_N=T$ of $N$ independent Brownian walkers
in one dimension.
Our main results have been obtained for two special geometries: \textit{(i)} Brownian motions (BM) starting from the same point $0$ and \textit{(ii)} Brownian bridges (BB). We have obtained explicit expressions for the moment generating function (MGF), i.e. the expectation of $e^{- c {\cal T}_N}$. We have mapped, through a Feynman-Ka\v c path integral representation, the determination of the MGF to the calculation of a Green function in the Lieb-Liniger model 
of quantum particles interacting with a pairwise delta interaction. Restricting to Brownians all starting all at $0$ allows
to consider only bosonic states, i.e. the delta Bose gas. For $c>0$ the MGF is the standard Laplace transform 
of $P_N(T)$, for which we obtained a formula for any $N$ using the eigenstate (spectral) decomposition
of the repulsive Bose gas. Laplace inversion then led us to
a compact formula for $P_N(T)$ for each geometry, one involving a determinant (for the BB) and 
the other one a Pfaffian (for the BM).\\

 We found that at small $T$ the PDF vanishes with two different sets of exponents
for the BM and the BB, and we obtained their exact amplitudes, related to Selberg integrals.  We have displayed very explicit formula for $N=2,3$ which we checked with excellent accuracy using extensive numerical simulations of Browian motions. We also obtained the mean and the variance of the coincidence time ${\cal T}_N$ as a function of $N$. At large $N$, the mean grows as $N^2$ and the variance as $N^3$. We then considered the double limit of large $N$ and large $T$, and, using a Coulomb gas approach, we showed the existence of a large deviation tail $P_N(T) \sim e^{- \mathcal{O}(N^2)}$  for $T \sim N^{3/2}$. 
Although we obtained explicit formula 
for small $T/N^{3/2}$, obtaining the full solution remains a challenge.
Furthermore, investigation of other possible regimes in this double large $T,N$ limit 
remain for future work.\\

We have shown that for $c<0$ the MGF is related to the exponential moments 
of the one dimensional Kardar-Parisi-Zhang (KPZ) equation, equivalently to the moments
of the directed polymer in a random potential. These moments are calculated using a summation
over the eigenstates of the attractive Lieb-Liniger model. These include bound states 
called strings. Here we obtained the large $T$ asymptotics of $P_N(T)$
from the contribution of the ground state of the Lieb-Liniger model, for the BB and the BM.
We were able to extend this asymptotics to arbitrary fixed initial and final
endpoints for the BM. Our main result is that the PDF of the coincidence time has a universal decay at
large $T$, of the form $P_N(T) \sim \exp(- 3 T^2/(N^3-N))$, and only the 
pre-exponential factor depends on the geometry. For the BB we used the connection to the droplet solution
of the KPZ equation, and for the BM to the flat initial condition.
It would be interesting to use other known solutions of the KPZ equation, such as the stationary initial condition
or KPZ in a half-space, to obtain properties on the coincidence time of Brownian walkers with different constraints. 
More generally, it would interesting to establish other universal properties of the distribution coincidence time 
using the knowledge of the spectral properties of the Lieb-Liniger model.
We hope that this work will motivate further studies of the coincidence properties of multiple diffusions.

\section*{Acknowledgements}
We are greatly indebted to G. Barraquand for numerous interactions during the preparation of this manuscript. B.LACT would like to thank D. Mavroyiannis for bringing this problem to his attention. We are also grateful to S.~N. Majumdar and G.~Schehr for interesting discussions. We acknowledge support from ANR grant ANR-17-CE30-0027-01 RaMaTraF.

\appendix
\section{Expected value of the coincidence time for non identical diffusing particles}\label{mean_var}

Considering independent particles with different diffusion coefficients $D_i$ and initial velocity $v_i$, we may compute the mean coincidence time using the independence together with the Brownian propagator
\be
P_i(x,t|0,0)=\frac{e^{-\frac{(x-v_i t)^2}{4D_i t}}}{\sqrt{4\pi D_i t}}\;.
\ee
We obtain for the mean coincidence time
\be
\moy{{\cal T}_N}=\sum_{i\neq j}\int_0^{1}\rmd \tau \int_{-\infty}^{\infty}\rmd x P_i(x,t|0,0)P_j(x,t|0,0)=\sum_{i\neq j}\frac{1}{v_i-v_j}\erf\left(\frac{v_i-v_j}{2\sqrt{D_i+D_j}}\right)\;.
\ee
We verify that this result only depends on the difference of speeds $v_i-v_j$ and not on the speed of the centre of mass $\bar v=\frac{1}{N} \sum_{i=1}^N v_i$. It is therefore not affected by a global shift of the speed $v_i\to v_i+v$.

\section{Details of the calculations of the distribution for $N=3$ Brownian bridges }\label{PDF_3_ap}

In this section, we detail the steps to obtain Eq. \eqref{P_3_BB_ana}. We start from Eq. \eqref{P_3_BB_nonex} that we reproduce here
\be\label{P_3_BB_nonex_ap}
P_{3,{\rm BB}}(T)=\partial_T^{4} \int_{0}^{T}\rmd r_1\int_{0}^{T-r_1}\rmd r_2\int_{0}^{T-r_1-r_2}\rmd r_3\left[1-3e^{-\frac{(r_1-r_{2})^2}{2}}+2\prod_{i=1}^3 e^{-\frac{(r_i-r_{i+1})^2}{4}}    \right]\;,
\ee
The first term in the integrand will give a zero contribution as after integration it is proportional to $T^3$ and derived four times with respect to $T$.
The second term can be computed as
\begin{align}
&-3\partial_T^4 \int_{0}^{\infty}\rmd u_1\int_{0}^{\infty}\rmd u_2\int_{0}^{\infty}\rmd u_3 \Theta\left(T-u_1-u_2-u_3\right)e^{-\frac{(u_1-u_2)^2}{2}}\nn\\
&=-3\partial_T^4 \int_{0}^{T}\rmd u_1\int_{0}^{T-u_1}\rmd u_2\left(T-u_1-u_2\right)e^{-\frac{(u_1-u_2)^2}{2}}\nn\\
&=-3\partial_T^3 \int_{0}^{T}\rmd u_1\int_{0}^{T-u_1}\rmd u_2 e^{-\frac{(u_1-u_2)^2}{2}}\nn\\
&=-3\partial_T^2 \int_{0}^{T}\rmd u_1 e^{-\frac{(2u_1-T)^2}{2}} =-3\partial_T^2 \int_{0}^{T}\rmd u_1 e^{-\frac{u_1^2}{2}}  \nn\\
&=-3\partial_T e^{-\frac{T^2}{2}}=3Te^{-\frac{T^2}{2}}\;.\label{second}
\end{align}
Finally, the third term in the integrand of Eq. \eqref{P_3_BB_nonex_ap} can also be obtained explicitly
\begin{align}
&2\partial_T^4 \int_{0}^{T}\rmd u_1\int_{0}^{T-u_1}\rmd u_2\int_{0}^{T-u_1-u_2}\rmd u_3  e^{-\frac{1}{4}\left[(u_1-u_2)^2+(u_2-u_3)^2+(u_1-u_3)^2\right]}\nn\\
&=2\partial_T^3 \int_{0}^{T}\rmd u_1\int_{0}^{T-u_1}\rmd u_2 e^{-\frac{1}{4}\left[(u_1-u_2)^2+(2u_2+u_1-T)^2+(2u_1+u_2-T)^2\right]}\nn\\
&=2\partial_T^3 \int_{0}^{T}\rmd u_1 e^{-\frac{\left(T-3u_1\right)^2}{8}}\int_{-\frac{1}{2}\left(T-u_1\right)}^{\frac{1}{2}\left(T-u_1\right)}\rmd u_2  e^{-\frac{1}{4}\left[\left(\frac{3}{2}u_1-\frac{T}{2}-u_2\right)^2+4u_2^2+\left(\frac{3}{2}u_1-\frac{T}{2}+u_2\right)^2\right]}\nn\\
&=2\partial_T^3 \int_{0}^{T}\rmd u_1 e^{-\frac{\left(T-3u_1\right)^2}{8}}\int_{-\frac{1}{2}\left(T-u_1\right)}^{\frac{1}{2}\left(T-u_1\right)}\rmd u_2 e^{-\frac{3}{2} u_2^2}\nn\\
&=4\partial_T^3 \int_{0}^{T}\rmd u_1 e^{-\frac{\left(T-3u_1\right)^2}{8}}\sqrt{\frac{\pi}{6}}\erf\left(\sqrt{\frac{3}{8}}\left(T-u_1\right)\right)\nn\\
&=4\partial_T^3 \int_{0}^{T}\rmd u_1 e^{-\frac{\left(2T-3u_1\right)^2}{8}}\sqrt{\frac{\pi}{6}}\erf\left(\sqrt{\frac{3}{8}}u_1\right)\nn\\
&=4\partial_T^2 \left[e^{-\frac{T^2}{8}}\sqrt{\frac{\pi}{6}}\erf\left(\sqrt{\frac{3}{8}}T\right)-\int_{0}^{T}\rmd u_1 \frac{1}{2}\left(2T-3u_1\right)e^{-\frac{\left(2T-3u_1\right)^2}{8}}\sqrt{\frac{\pi}{6}}\erf\left(\sqrt{\frac{3}{8}}u_1\right)\right]\nn\\
&=4\partial_T^2 \left[e^{-\frac{T^2}{8}}\sqrt{\frac{\pi}{6}}\erf\left(\sqrt{\frac{3}{8}}T\right)-\frac{2}{3}\sqrt{\frac{\pi}{6}}e^{-\frac{T^2}{8}}\erf\left(\sqrt{\frac{3}{8}}T\right)+\frac{1}{3}\int_{0}^{T}\rmd u_1 e^{-\frac{\left(2T-3u_1\right)^2+3 u1^2}{8}}\right]\nn\\
&=4\partial_T^2 \left[\frac{1}{3}\sqrt{\frac{\pi}{6}}e^{-\frac{T^2}{8}}\erf\left(\sqrt{\frac{3}{8}}T\right)+\frac{1}{3}e^{-\frac{T^2}{8}}\int_{-\frac{T}{2}}^{\frac{T}{2}}\rmd u e^{-\frac{3}{2}u^2}\right]\nn\\
&=4\partial_T^2 \left[\sqrt{\frac{\pi}{6}}e^{-\frac{T^2}{8}}\erf\left(\sqrt{\frac{3}{8}}T\right)\right]\label{third}
\end{align}
Computing the derivative in Eq. \eqref{third} and combining this term with Eq. \eqref{second}, we obtain the final result for $N=3$ Brownian bridges in Eq. \eqref{P_3_BB_ana}\;.

\section{Details of the numerical simulations}\label{simul}

Our numerical simulations of the Brownian motion are realized from independent random walks
\be
x_{{\rm BM},i+1,j}=x_{{\rm BM},i,j}+\sqrt{2}\eta_{i+1,j}\;,\;\;x_0=0\;,
\ee
for $T=10^4$ steps and $j=1,\cdots,N$ where the $\eta_{i,j}$'s are Gaussian centered i.i.d random variables of unit variance. To simulate the Brownian bridge, we consider the process
\be
x_{{\rm BB},i,j}=x_{{\rm BM},i,j}-\frac{i}{T}x_{{\rm BM},T,j}\;,
\ee
to ensure $x_{{\rm BB},T}=x_{{\rm BB},0}=0$. To measure the local time we just add for each $j\neq k$ the number of steps $i$ where $|x_{i,j}-x_{i,k}|<\ell$ and divide by $2\ell T$. We chose for the simulations $\ell=0,01$.

{}

\end{document}